\documentclass[useAMS,usenatbib]{mnras}
\usepackage{graphicx,natbib,amssymb,amsmath,stmaryrd,makecell}
\usepackage{soul}
\usepackage{float}
\usepackage{dblfloatfix}
\usepackage{subcaption}
\begin{document}
\title{Dark Energy Constraints from the Thermal Sunyaev Zeldovich Power Spectrum}
\author[B. Bolliet et al.]{Boris Bolliet,$^{1,4}$  Barbara Comis,$^1$
Eiichiro Komatsu,$^{2,3}$ Juan Francisco Mac\'{\i}as-P\'erez$^1$\\
$^1$Laboratoire de Physique Subatomique et de Cosmologie, Universit\'e
Grenoble-Alpes,\\ CNRS/IN2P3, 53, 
avenue des Martyrs, 38026 Grenoble
cedex, France\\
$^2$Max-Planck-Institut f{\"u}r Astrophysik, Karl-Schwarzschild Strasse
1, 85748 Garching, Germany\\
$^3$Kavli Institute for the Physics and Mathematics of the Universe
(Kavli IPMU, WPI), Todai Institutes for Advanced Study, \\ University
of Tokyo, Kashiwa 277-8583, Japan\\
$^4$Jodrell Bank Centre for Astrophysics, School of Physics and Astronomy,
The University of Manchester, Manchester, M13 9PL, U.K.
}
\maketitle		
\begin{abstract}
  We constrain the dark energy equation of state parameter, $w$, using
  the power spectrum of the thermal Sunyaev-Zeldovich (tSZ) effect. We
  improve upon previous analyses by taking into account the trispectrum
  in the covariance matrix and marginalising over the foreground
  parameters, the correlated noise, the mass bias $B$ in the Planck
  universal pressure profile, and all the relevant cosmological
  parameters (i.e., not just $\Omega_{\mathrm{m}}$ and $\sigma_8$). We
  find that the amplitude of the tSZ power spectrum at $\ell\lesssim
  10^3$ depends primarily on $F\equiv
  \sigma_{8}(\Omega_{{\mathrm{m}}}/B)^{0.40}h^{-0.21}$, where $B$ is
  related to more commonly used variable $b$ by $B=(1-b)^{-1}$. We
 measure this parameter with 2.6\% precision, $F=0.460\pm 0.012$
 (68\%~CL). By fixing the bias to $B=1.25$ and adding the local determination of the Hubble constant $H_0$ and the amplitude of the primordial power spectrum constrained
  by the Planck Cosmic Microwave Background (CMB) data, we find
  $w=-1.10\pm0.12$, $\sigma_{\mathrm{8}}=0.802\pm0.037$, and
  $\Omega_{{\mathrm{m}}}=0.265\pm0.022$ (68\%~CL). Our limit on $w$ is
 consistent with and is as tight as that from the distance-alone constraint
  from the CMB and $H_0$. Finally, by combining the tSZ power spectrum
  and the CMB data we find, in the $\Lambda$ Cold Dark Matter (CDM)
  model, the mass bias of $B=1.71\pm 0.17$, i.e., $1-b=0.58\pm 0.06$
  (68\%~CL).
\end{abstract}
\section{Introduction}\label{sec:IN}
Clusters of galaxies are the largest gravitationally bound objects in the universe and constitute compelling probes for cosmological studies. 
In particular, the angular power spectrum of the thermal Sunyaev Zeldovich (tSZ) effect
\citep{1972CoASP...4..173S} depends sensitively on the amplitude of
matter fluctuations \citep{Komatsu:1999ev,Komatsu:2002wc}. While at large
multipoles ($\ell\gtrsim 10^3$) the power spectrum depends also on
the details of pressure profiles within halos, at smaller multipoles the
dependence is much weaker \citep{Komatsu:1999ev,McCarthy:2013qva}. This
makes the tSZ power spectrum at $\ell\lesssim 10^3$ a powerful probe of cosmology.

Dark energy slows down structure formation \citep[see][for a recent
review]{Weinberg:2012es}. A less negative value of the dark energy
equation of state (EoS) parameter, $w$,
makes dark energy dominate at higher redshifts, suppressing the growth
of structure, hence the present-day matter amplitude parameter
$\sigma_8$. Using this anti-correlation between $w$ and $\sigma_8$
\citep{Komatsu:2008hk}, we can constrain the nature of dark energy.
We achieve this by comparing the amplitude of matter
fluctuations in a late time universe with that at the last scattering
surface of the cosmic microwave background (CMB). 

The Planck Collaboration delivered an all-sky map of the Compton $y$
parameter as well as an estimate of the angular power spectrum up to
$\ell\simeq 1300$ \citep{Ade:2013qta,Aghanim:2015eva}. They obtained a
constraint on the parameter combination
$\sigma_8\Omega_{\mathrm{m}}^{3/8}$, with the nuisance parameters such
as the foreground  contaminants and residual correlated noise marginalised over, but with the
mass bias and all the other cosmological parameters fixed, and with the power spectrum 
covariance matrix containing only a Gaussian term. 
\begin{table*}
\begin{centering}
\setcellgapes{2pt}\makegapedcells
\begin{tabular}{l|cccccccccc}
 &$A_0$ & $a_0$ & $b_0$ & $c_0$&$A_z$ & $a_z$ & $b_z$ & $c_z$&&\tabularnewline
\hline 
\cite{Bocquet:2015pva}   & 0.228 & 2.15 & 1.69 &1.30 & 0.285 & -0.058 & -0.366 & -0.045&&\tabularnewline
\cite{Tinker:2008ff}   & 0.186 & 1.47 & 2.57 &1.19 & -0.14 & -0.06 & -0.011 & 0&&\tabularnewline
\hline
\tabularnewline
 &$\alpha_0$ & $\beta_0$ & $\gamma_0$ & $\eta_0$&$\phi_0$ & $\alpha_z$ & $\beta_z$ & $\gamma_z$&$\eta_z$&$\phi_z$\tabularnewline
 \hline
\cite{2010ApJ...724..878T}  & 0.368 & 0.589 & 0.864 & -0.243 & -0.729 & 0 & 0.2 & -0.01&0.27&-0.08
\end{tabular}
\par\end{centering}
\caption{Parameters for the halo mass functions (HMF). Note that these parameters values are relevant for \protect\cite{Bocquet:2015pva}, \protect\cite{Tinker:2008ff} and  \protect\cite{2010ApJ...724..878T}  HMFs evaluated at the over-density mass $M_{200m}$ (for the \protect\cite{Tinker:2008ff} formula at $M_{1600m}$, the value of $b_z$ has to be replaced by  $b_z=-0.314$). Given a parameter $p=A,b,..,$ the redshift dependence is obtained as $p=p_0 (1+z)^{p_z}$.\label{tab:HMFparams}}
\end{table*}
\cite{Horowitz:2016dwk} revisited the Planck analysis by including the trispectrum term in the covariance,
but held the nuisance parameters fixed at the best-fitting values
obtained by the Planck Collaboration.
\cite{Hurier:2017jgi} varied the mass bias, included the trispectrum in
the covariance but used a different method for the tSZ power spectrum estimation. \cite{Salvati:2017rsn}
included the trispectrum and varied the mass bias and all the relevant
cosmological parameters, but not the amplitude of the foreground contaminants and correlated noise.

In this paper, we obtain constraints on the parameter combination that
determines the amplitude of the tSZ power spectrum at $\ell\lesssim
10^3$ and $w$ by including the trispectrum in the covariance, as well as
by varying the
mass bias, the nuisance parameters, and all the relevant cosmological
parameters. 

The rest of the paper is organized as follows.
In section \ref{sec:AnalyticalCalc} we describe main steps of our
calculation of the tSZ power spectrum and its numerical
implementation. In section~\ref{sec:what?} we find the parameter
combination that scales the amplitude of the tSZ power.
In section \ref{sec:MLA} we present settings of our
likelihood analysis and show importance of the non-Gaussian contribution
to the covariance matrix. In section \ref{sec:Results} we present our
cosmological constraints. We conclude in section \ref{sec:SC}.

\section{Model for the tSZ power spectrum}\label{sec:AnalyticalCalc}
Our model consists of the halo mass function (HMF) and the pressure
profile of the intra-cluster medium (ICM). We consider only the $1$-halo
contribution, as the $2$-halo term contribution to the tSZ power
spectrum is not significant compared to precision of the current data
\citep{Komatsu:1999ev}.

For numerical calculations of the tSZ power spectrum (and trispectrum),
we have developed a version of the publicly-available package \verb|class|
\citep{2011arXiv1104.2932L,2011JCAP...07..034B} augmented with a tSZ module
in C. The code is dubbed \verb|class_sz| and is available on the
internet\footnote{website: \href{https://github.com/borisbolliet/class_sz_public}{https://github.com/borisbolliet/class\_sz\_public}}. Our code
builds upon and improves performance of the previous code \verb|szfast|
in Fortran 90 \citep{Komatsu:2002wc,Dolag:2015dta}.

The tSZ angular power spectrum is calculated via 
\begin{equation}
C_{\ell}^{{\mathrm{tSZ}}}= \int_{z_{{\mathrm{min}}}}^{z_{{\mathrm{max}}}}\mathrm{d}z\frac{dV}{dzd\Omega}\int_{\ln M_{\mathrm{min}}}^{\ln M_{\mathrm{max}}}d\ln M\frac{dn}{d\ln M}\left|y_{\ell}\left(M,z\right)\right|^{2},\label{eq:cltSZ-1}
\end{equation}
where $y_\ell$ is the two dimensional Fourier transform of an electron
pressure profile, $dn/dM$ is the HMF, and $M$ is a characteristic mass
of dark matter halos which will be defined more precisely later. The integration
over mass is performed using a Gaussian quadrature method within
the mass range of $M_{{\mathrm{min}}}=10^{11}h^{-1}\mathrm{M}_{\varodot}\,\,\,\,\mathrm{and}\,\,\,\,\,M_{{\mathrm{max}}}=5\times10^{15}h^{-1}\mathrm{M}_{\varodot}$, where $h\equiv H_{0}/100$ is the reduced Hubble constant and $\mathrm{M}_{\varodot}$ is the solar mass. This mass range is chosen so that the integral over the mass converges.
In Eq.~\eqref{eq:cltSZ-1}, $V$ is the comoving volume of the universe and its derivative with respect to redshift $z$ and solid angle $\Omega$ is given by
\begin{equation}
\frac{dV}{dzd\Omega}=\left(1+z\right)^{2}d^2_{{\mathrm{A}}}(z)c/H(z),
\end{equation}
where  $c$ is the speed of light, $d_{{\mathrm{A}}}\left(z\right)$ is the
physical angular diameter distance, and $H(z)$ the Hubble
expansion rate.

The integration over redshift is carried out with a simple
trapezoidal rule from $z_{{\mathrm{min}}}=0$ and
up to $z_{{\mathrm{max}}}=6$. At higher redshift
the number density of halos is vanishingly small.
 
Using Limber's approximation, the two dimensional Fourier transform of an electron pressure profile
$P_e$ is given by \citep[see, e.g.,][]{Komatsu:2002wc}
\begin{equation}\label{eq:yl}
y_{\ell} =\frac{\sigma_{{\scriptscriptstyle
 \mathrm{T}}}}{m_{{\mathrm{e}}}c^{2}}\frac{4\pi
 r_{{\mathrm{500}}}}{\ell_{{\scriptscriptstyle
 \mathrm{500}}}^{2}}\int_{x_{\rm min}}^{x_{\rm max}}\mathrm{d}xx^{2}\frac{\sin\left(\ell x/\ell_{{\mathrm{500}}}\right)}{\ell x/\ell_{{\mathrm{500}}}}P_{{\mathrm{e}}}\left(x\right),
\end{equation}
where $\sigma_{{\scriptscriptstyle
 \mathrm{T}}}$ is the Thomson scattering cross section, $m_{{\mathrm{e}}}$ is the electron mass, $x\equiv r/r_{{\mathrm{500}}}$ with $r$ being
the radial distance to the center of the halo, $r_{{\scriptscriptstyle
\mathrm{500}}}$ the radius of a sphere containing the over-density mass
$M_{{\mathrm{500c}}}$ of 500 times the critical
density of the universe,  and  $\ell_{{\scriptscriptstyle
\mathrm{500}}}\equiv d_{{\scriptscriptstyle
\mathrm{A}}}/r_{{\mathrm{500}}}$.
The integral in Eq.~\eqref{eq:yl} is performed with Romberg's
method between $x_{\rm min}=10^{-6}$ and $x_{\rm max}=10$. 

For the pressure profile, we use a standard generalized Navarro-Frenk-White parametrization:
\begin{equation}\label{eq:pf}
P_{{\mathrm{e}}}\left(x\right)=C\times P_{0}\left(c_{{\mathrm{500}}}x\right)^{-\gamma}\left[1+\left(c_{{\mathrm{500}}}x\right)^{\alpha}\right]^{\left(\gamma-\beta\right)/\alpha},
\end{equation}
where $\left\{ \gamma,\alpha,\beta,P_{0},c_{{\mathrm{500}}}\right\} $
are set to their best-fitting values obtained by \cite{2013A&A...550A.131P}.
The reader is referred to Appendix D of \cite{2011ApJS..192...18K}
for further details regarding this parameterization.
We do not vary these parameters in this paper, which allows us to speed
up our likelihood analysis by tabulating $y_\ell$. The coefficient $C$ depends on mass as
\begin{equation}
C=1.65~\left(\frac{h}{0.7}\right)^{2}\left(\frac{H}{H_{0}}\right)^{\frac{8}{3}}\left[\frac{(h/0.7)\tilde M_{{\mathrm{500c}}}}{3\times10^{14}\mathrm{M}_{\varodot}}\right]^{\frac{2}{3}+0.12}\,\,\mathrm{eV}\,\mathrm{cm}^{-3}.\label{eq:pressureprofile-1}
\end{equation}
The mass used in Eq.~\eqref{eq:pressureprofile-1}, $\tilde
M_{{\mathrm{500c}}}$, is not necessarily the true mass but may contain a
bias due to non-thermal pressure, observational effects, etc. To
account for a possible bias, we relate the true mass to $\tilde
M_{{\mathrm{500c}}}$ as $\tilde{M}_{{\mathrm{500c}}}=M_{{\mathrm{500c}}}/B$. In
the literature, a different variable $b$ is often used \citep{Ade:2013qta}, and it is
related to $B$ via $B=(1-b)^{-1}$.
Note that this rescaling not only affects the normalisation of the
pressure profile but also its scale dependence via
$\ell_{{\mathrm{500}}}$ because it modifies $r_{\mathrm{500}}\propto M_{\mathrm{500}}^{1/3}$.

The HMF in Eq. \eqref{eq:cltSZ-1} depends on both mass and redshift and is written as 
\begin{align}
\frac{dn}{d\ln M} & =-\tfrac{1}{2}f\left(\sigma,z\right)\frac{\rho_{{\mathrm{m}}0}}{M}\frac{d\ln\sigma^{2}}{d\ln M}\,,
\end{align}
where $\rho_{{\mathrm{m}}0}$ is the present-day mean mass density of the
Universe, $\sigma^{2}$ is the variance of the matter over-density field
smoothed by a sphere of radius $R\equiv\left[3M/4\pi\rho_{{\mathrm{m}}0}\right]^{1/3}$,
i.e.,
\begin{equation}
\sigma^{2}\left(M,z\right)\equiv\int_{k_{\mathrm{min}}}^{k_{\mathrm{max}}}\frac{dk}{k}\frac{k^{3}}{2\pi^{2}}P\left(k,z\right)W^2\left(kR\right)\label{eq:sigma2m}\,,
\end{equation}
where $P\left(k\right)$ is the linear matter power spectrum, with $k_{\mathrm{min}}=10^{-4}~h\mathrm{Mpc^{-1}}$ and $k_{\mathrm{max}}=50~h\mathrm{Mpc^{-1}}$, 
$W$ is the three dimensional Fourier transform of a top-hat window
function and $f\left(\sigma,z\right)$ is often measured from N-body
simulations. We shall specify the form of $f\left(\sigma,z\right)$
later.

Since the HMF is generally parameterized in terms of the over-density
mass $M_{\mathrm{X}}$, we write
\begin{equation}
\frac{dn}{d\ln M} =\frac{d\ln M_{\mathrm{X}}}{d\ln M}\frac{dn}{d\ln M_{\mathrm{X}}}=-\frac{d\ln M_{\mathrm{X}}}{d\ln M}\frac{1}{8\pi R^{3}_{\mathrm{X}}}\frac{d\ln\sigma^{2}}{d\ln R_{\mathrm{X}}}f\left(\sigma,z\right) \label{eq:dndlnM} \nonumber
 \end{equation}
 with $R_{\mathrm{X}}\equiv\left[3M_{\mathrm{X}}/4\pi\rho_{{\mathrm{m}}0}\right]^{1/3}$,
and use an approximation $d\ln M_{\mathrm{X}}/d\ln M\approx1$
\citep[see, e.g.,][]{Komatsu:2001dn}. \\
\begin{figure}
\begin{centering}
\includegraphics[width=7cm]{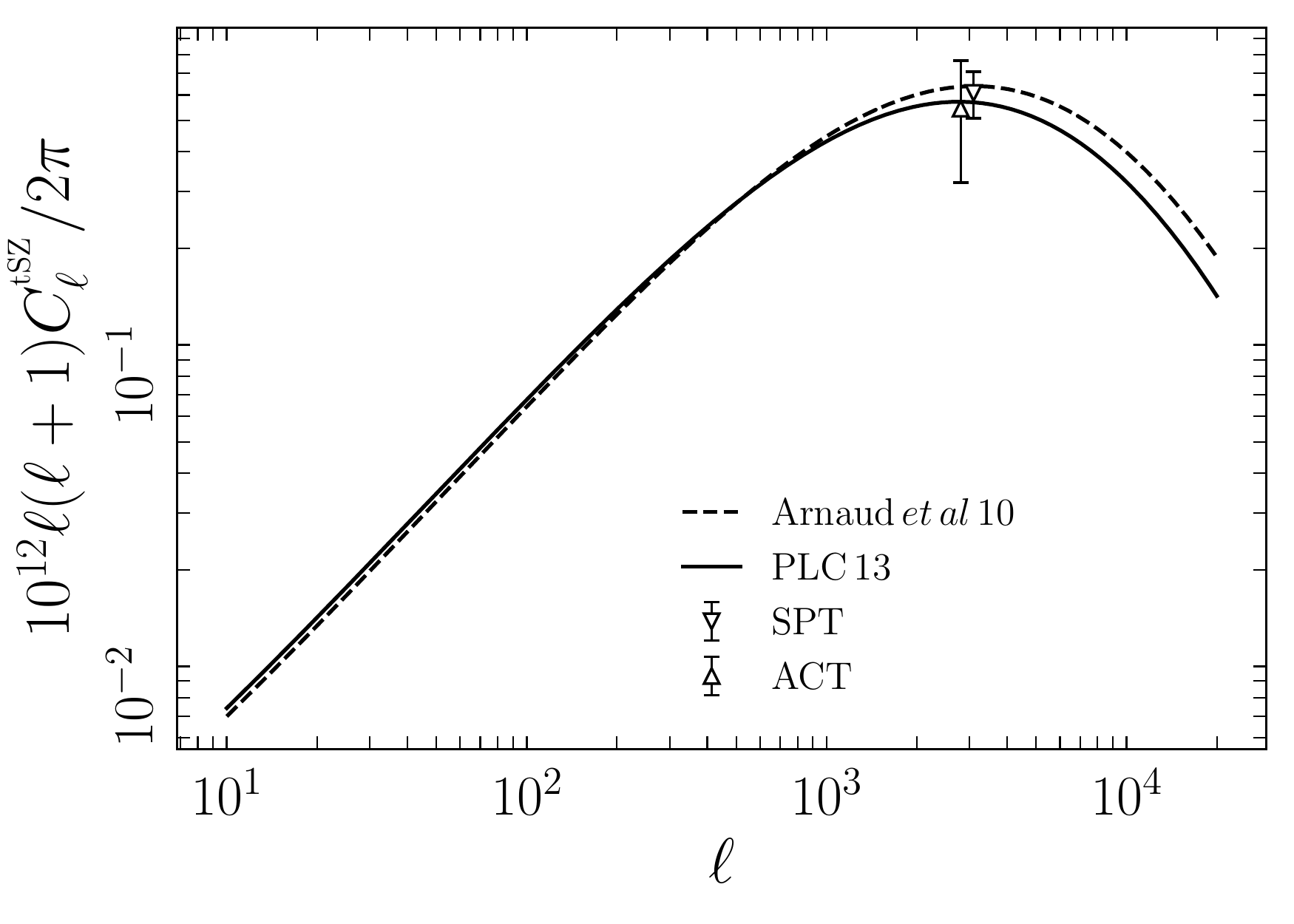}
\includegraphics[width=7cm]{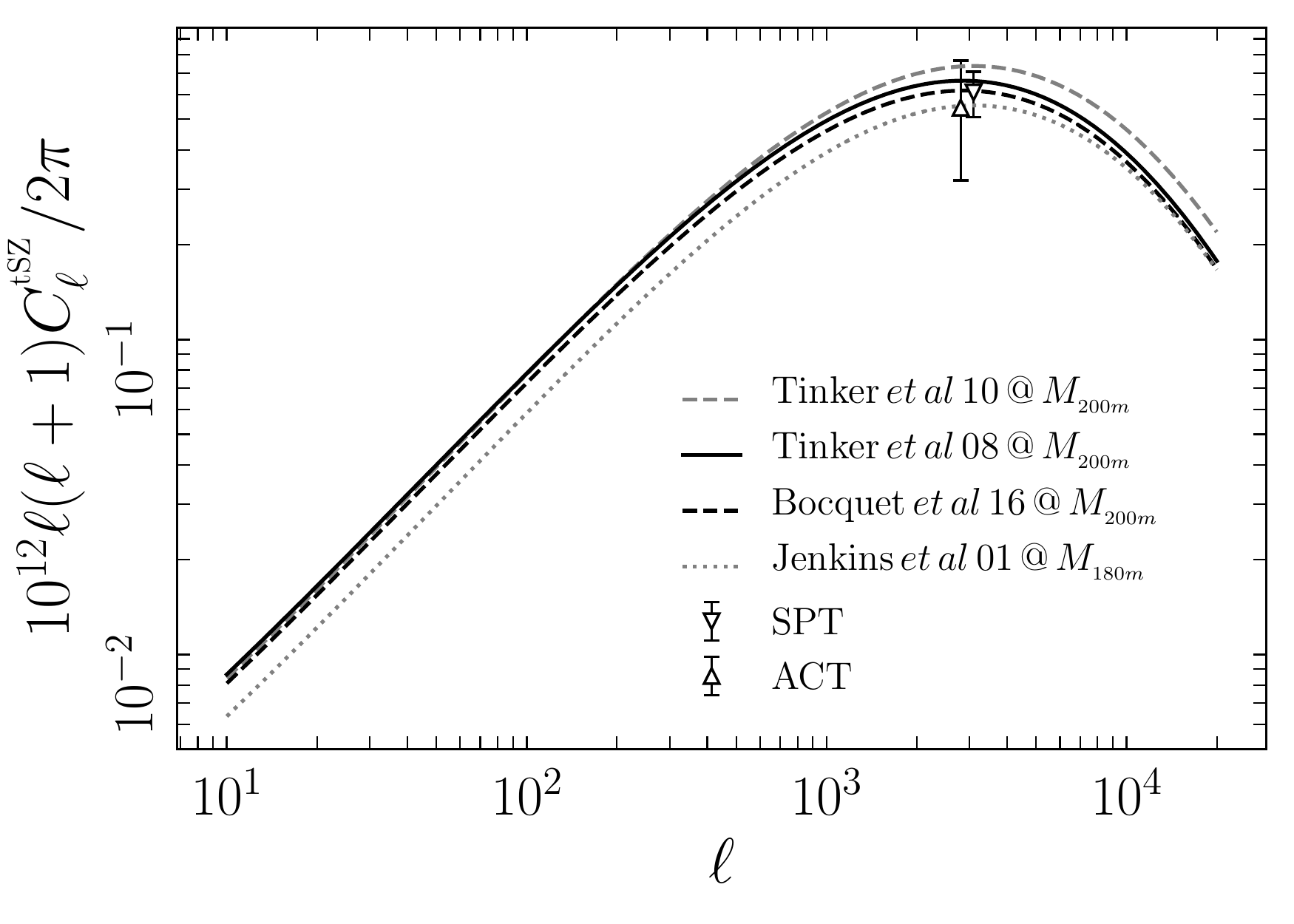}
\includegraphics[width=7cm]{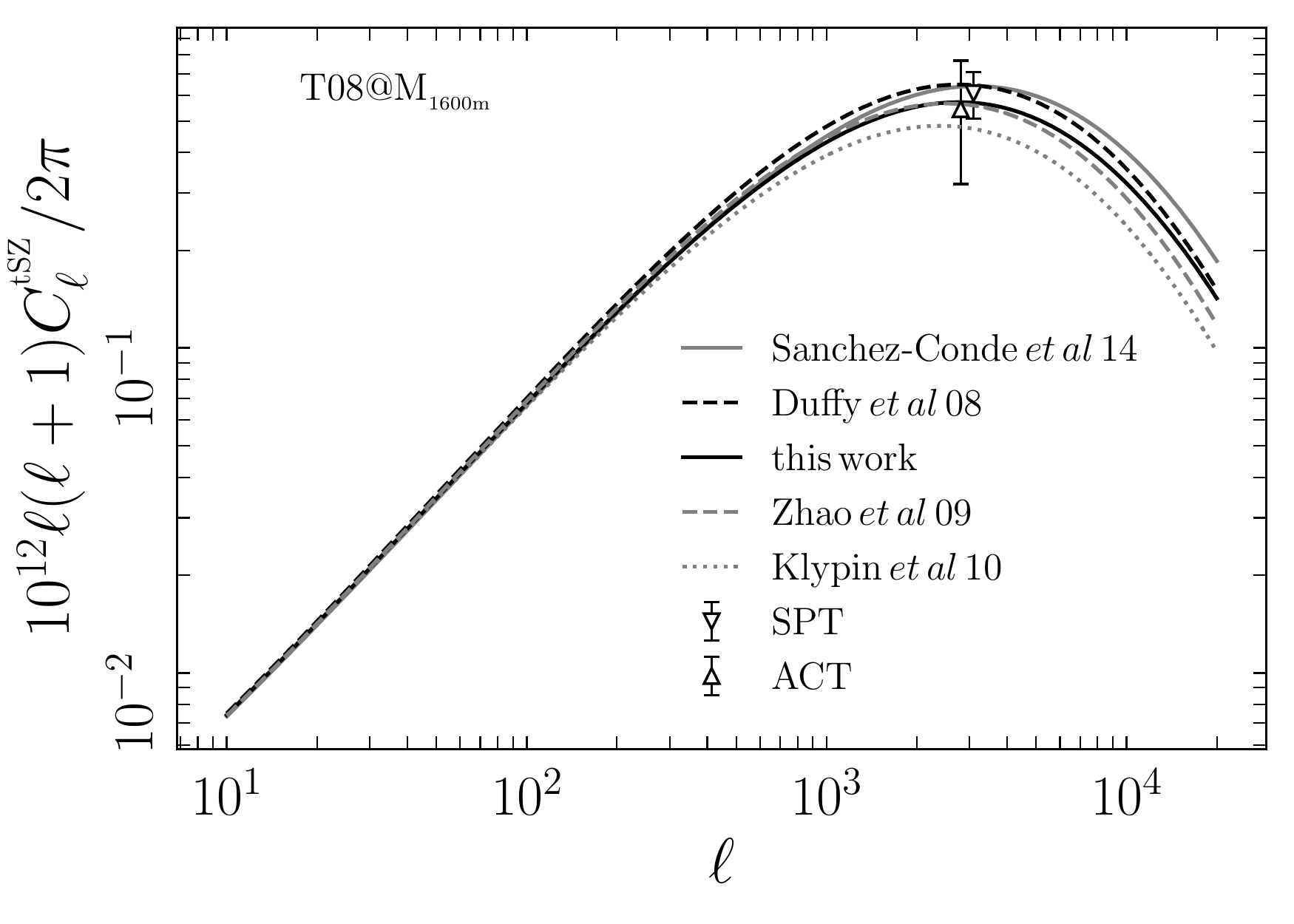}
\includegraphics[width=7cm]{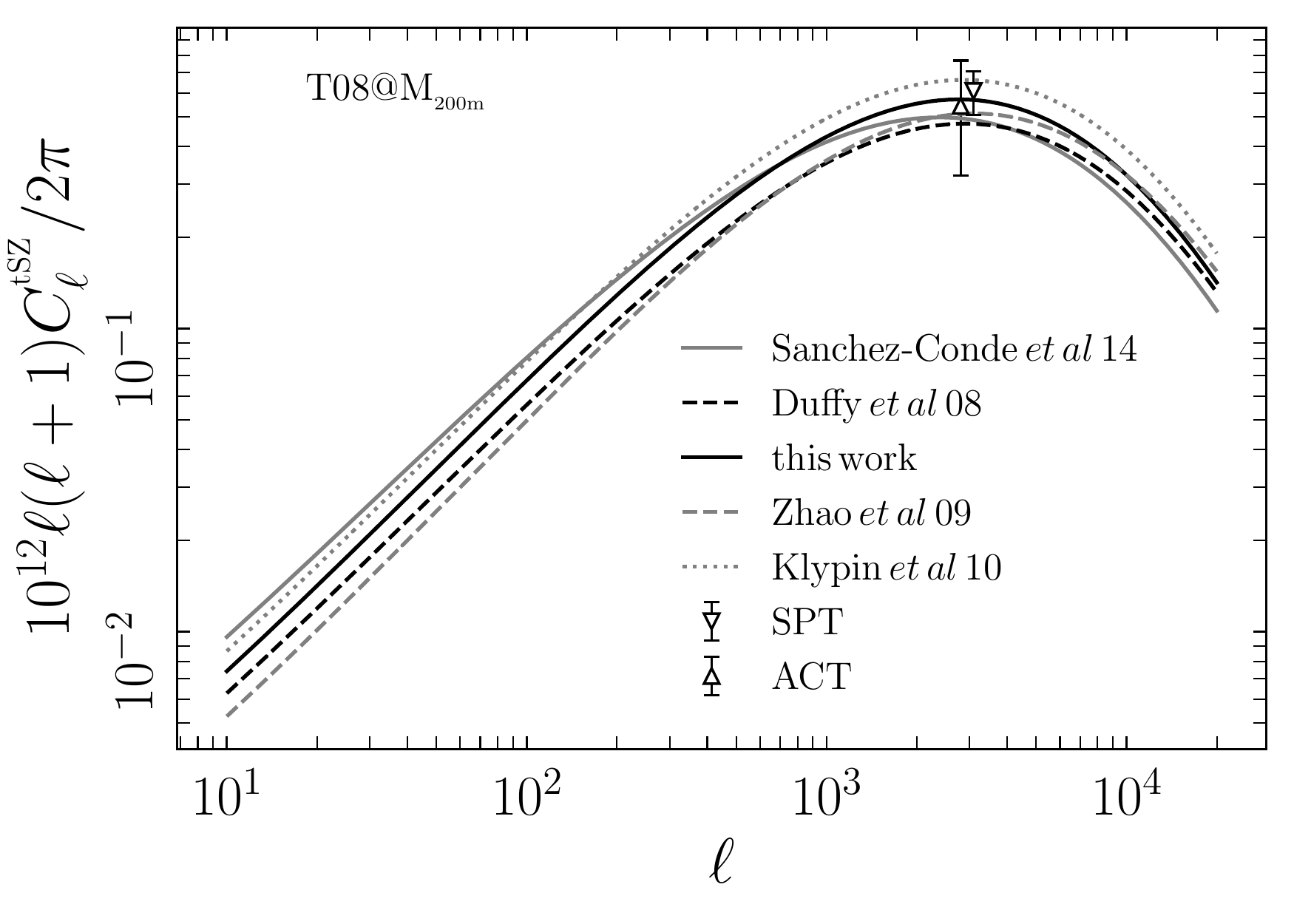}
\par\end{centering}
 \caption{tSZ power spectra computed with various settings. From top to
 bottom panels we show two pressure profiles; four HMFs evaluated at
 $M_{\mathrm{200m}}$ or $M_{\mathrm{180m}}$ with mass conversion done by
 the Klypin et al. 2010 concentration-mass relation; the \protect\cite{Tinker:2008ff} HMF (T08) evaluated at
 $M_{\mathrm{1600m}}$ with mass conversion done by four concentration-mass
 relations; and same but for
 $M_{\mathrm{200m}}$. The solid black lines in the bottom two panels are
 computed with the \protect\cite{Tinker:2008ff} HMF evaluated at $M_{\mathrm{500c}}$, so that no
 mass conversion is needed.  The fiducial model is $\Lambda$CDM with
 $\sigma_8 = 0.79$, $\Omega_{\mathrm{m}}=0.32$, $h=0.66$, $B=1.53$ and
 $n_{\mathrm{s}}=0.81$. The Atacama Cosmology Telescope \citep[ACT;][]{Hasselfield:2013wf} and South
 Pole Telescope  \citep[SPT;][]{George:2014oba} data points are shown for comparison.
 \label{fig:VaryingHMF}}
\end{figure}

We have implemented four different parameterisations
of the HMF: the \cite{Bocquet:2015pva}  fitting formula obtained
from the \verb|Magneticum| simulation with the impact of baryons; the \cite{Tinker:2008ff} formula; 
the \cite{2010ApJ...724..878T} formula, an updated version of the former;
the \cite{Jenkins:2000bv} formula. In particular, the \cite{Bocquet:2015pva} and \cite{Tinker:2008ff} HMFs are expressed as 
\begin{equation}
f\left(\sigma,z\right)=  A\left[\left(\frac{\sigma}{b}\right)^{-a}+1\right]\exp\left(-\frac{c}{\sigma^{2}}\right).\label{eq:MF}
\end{equation}
The  \cite{2010ApJ...724..878T}  HMF is parameterized as
\begin{equation}
f\left(\nu,z\right)=  \alpha\left[\left(\beta^2 \nu\right)^{-\phi}+1\right]\nu^{\eta}\exp\left(-\gamma\frac{\nu}{2}\right)\sqrt{\nu},\label{eq:MF-T10}
\end{equation}
where $\nu$ is defined via $\sigma=1.685/\sqrt{\nu}$. The fitting parameters of these HMFs depend on redshift and are reported in table \ref{tab:HMFparams}. The \cite{Jenkins:2000bv} formula for the HMF evaluated at  $M_{\mathrm{180m}}$ (over-density mass of 180 times the mean matter density) does not have an explicit redshift dependence and reads as
$f\left(\sigma\right)= 0.301\exp\left(-|0.64-\ln\sigma|^{3.82}\right)$.\\

The HMF is often given for various over-density masses
$M_{\mathrm{X}}$. This could be, e.g.,  $M_{\mathrm{200m}}$
(over-density mass of 200 times the mean matter density),
$M_{\mathrm{180m}}$ or $M_{\mathrm{500c}}$. There have been two approaches to treat the differences in the mass
definitions: 
\begin{enumerate}
  \item \cite{Aghanim:2015eva,Salvati:2017rsn,
	Hurier:2017jgi} use $M_{\mathrm{500c}}$ in the integral of
	Eq.~\eqref{eq:cltSZ-1}. Then, no conversion is needed to compute
	the pressure profile which often takes $M_{\mathrm{500c}}$ as an
	input. For the HMF they used \cite{Tinker:2008ff} which provides fitting formulae
	for various over-density masses with respect to the mean mass
	density, but not for the critical over-density mass $M_{\mathrm{500c}}$; thus, using tables 2 and B3 of the reference they
	interpolated the HMF parameters at  $M_{\mathrm{500c}}$ at every redshift.\\
 \item The papers that follow \cite{Komatsu:2002wc} use the virial mass
       $M_{\mathrm{vir}}$ in the integral of
	Eq.~\eqref{eq:cltSZ-1}. Conversion from the virial mass to the
       over-density mass is needed twice: for the HMF and for the
       pressure profile. The conversion is carried out using the so-called 
       concentration-mass relation for dark matter halos. We have implemented four concentration-mass relations, including \cite{Duffy:2008pz}, i.e., 
       $c_{\mathrm{vir}} = 7.85\times(M_{\mathrm{vir}}/2\times10^{12})^{-0.81}\times(1+z)^{-0.71}$,
       where $M_{\mathrm{vir}}$ is in units $h^{-1}\mathrm{M_\odot}$; \cite{2011ApJ...740..102K}, i.e., 
       \begin{equation}
       c_{\mathrm{vir}} = c_0\times(M_{\mathrm{vir}}/10^{12})^{-0.075}\times[1+(M_{\mathrm{vir}}/M_{\mathrm{0}})^{-0.26}],\nonumber
       \end{equation}
       where $c_0$ and $M_0$ are functions of redshift and whose tabulation can be found in table 3 of the reference; \cite{Sanchez-Conde:2013yxa}, which uses concentrations  at the over-density mass
$M_{{\mathrm{200c}}}$ of 200 times the critical
density of the universe instead of $M_{\mathrm{vir}}$, i.e., 
       $c_{\mathrm{200}} = \sum_{i=0}^{5} c_i \times [\ln(M_{\mathrm{200c}})]^i \times(1+z)^{-1}$,
       where the values for the coefficients $c_i$ are given bellow Eq. 1 of the reference; and \cite{Zhao:2008wd} which does not give an explicit concentration-mass relation but computes it numerically at every redshift with the \verb|mandc| code\footnote{website: \href{http://202.127.29.4/dhzhao/mandc.html}{http://202.127.29.4/dhzhao/mandc.html}}. In this case, we ran  \verb|mandc| for the Planck 2015 best-fitting cosmological parameters and for several redshift values and tabulated the concentration for susequent interpolation. 
  \end{enumerate}
The original motivation behind the second approach was that the HMF was
thought to be more universal (i.e., the function $f(\sigma,z)$ is a
function of $\sigma$ only, without explicit dependence on $z$) when
using $M_{200\mathrm{m}}$. However,  \cite{Tinker:2008ff}  showed that the HMF is not
universal for any masses. This then motivates the first approach, which
calibrates the HMF directly for the relevant mass definition for the
pressure profile, such as $M_{500\mathrm{c}}$.
These two strategies have led to different conclusions regarding the
cosmological parameter constraints obtained from the SZ power spectrum
data. In particular,  the work based on the second approach has found a
higher value of $\sigma_{8}$ than the Planck Collaboration that uses the
first approach, by about two standard deviations. Here we show that this
discrepancy is not due to the use of different HMFs, but to the
ambiguity of the use of the concentration-mass relation.

In figure \ref{fig:VaryingHMF} we show the tSZ power spectra computed
with different settings. In the top panel we compare the power
spectra using the Planck pressure profile and that of the original work
by \cite{2010A&A...517A..92A}. They make little difference at
$\ell\lesssim 10^3$. In the second panel we compare the spectra using
the four halo mass functions (HMFs). While the \cite{Jenkins:2000bv} HMF gives somewhat lower
amplitude, the others give similar results. In the third panel we
compare the spectra using the \cite{Tinker:2008ff} HMF for $M_{\mathrm{1600m}}$ (which is
close to $M_{\mathrm{500c}}$) and the four concentration-mass 
relations. We find that all concentration-mass relations give similar results at $\ell\lesssim 10^3$
because conversion from $M_{500\mathrm{c}}$ to $M_{\mathrm{1600m}}$ is
small. However, in the last panel the four concentration-mass relations give diverging
results because we use the \cite{Tinker:2008ff} HMF for $M_{\mathrm{200m}}$. In
particular, the \cite{Duffy:2008pz} concentration-mass relation leads to an underestimation of the tSZ
power compared to the third panel. 
So, to fit the same data,
 tSZ power spectrum models with \cite{Duffy:2008pz}
 concentration-mass relation and $M_{\mathrm{200m}}$ need a larger $\sigma_8$ than models without mass conversion. In other words, using the approach of \cite{Komatsu:2002wc} with the \cite{Duffy:2008pz} concentration-mass relation for the mass conversion would lead to an
overestimation of $\sigma_8$.
\begin{table}
\begin{centering}
\setcellgapes{2pt}\makegapedcells
\begin{tabular}{c|cccc|cccc|cc}
 Parameter & Min. & Max. &  \tabularnewline
\hline
$10^{9}A_{{\mathrm{s}}}$ & 1.8 & 2.7 \tabularnewline
$n_{{\mathrm{s}}}$ & 0.8 & 1  \tabularnewline
$\tau_{{\mathrm{reio}}}$ & 0.04 & 0.12  \tabularnewline
$100\theta_{s}$ & 1.03 & 1.05 \tabularnewline
 $\Omega_{{b}}h^{2}$ & 0.0199 & 0.0245 \tabularnewline
 $\Omega_{{c}}h^{2}$ & 0.09 & 0.15 \tabularnewline
 $w$ & $-2$ & $-0.5$ \tabularnewline
 $B$ & 1.11 & 1.67 \tabularnewline
 $A_{{\mathrm{CIB}}}$ & 0 & 10 \tabularnewline
 $A_{{\mathrm{IR}}}$ & 0 & 10 \tabularnewline
 $A_{{\mathrm{RS}}}$ & 0 & 10 \tabularnewline
\end{tabular}
\par\end{centering}
\caption{Uniform priors on parameters of the MCMC analysis.\label{tab:Uniform-priors-imposed}}
\end{table}
\section{What determines the amplitude of the power spectrum?}
\label{sec:what?}
\begin{figure*}
 \includegraphics[width=5.7cm]{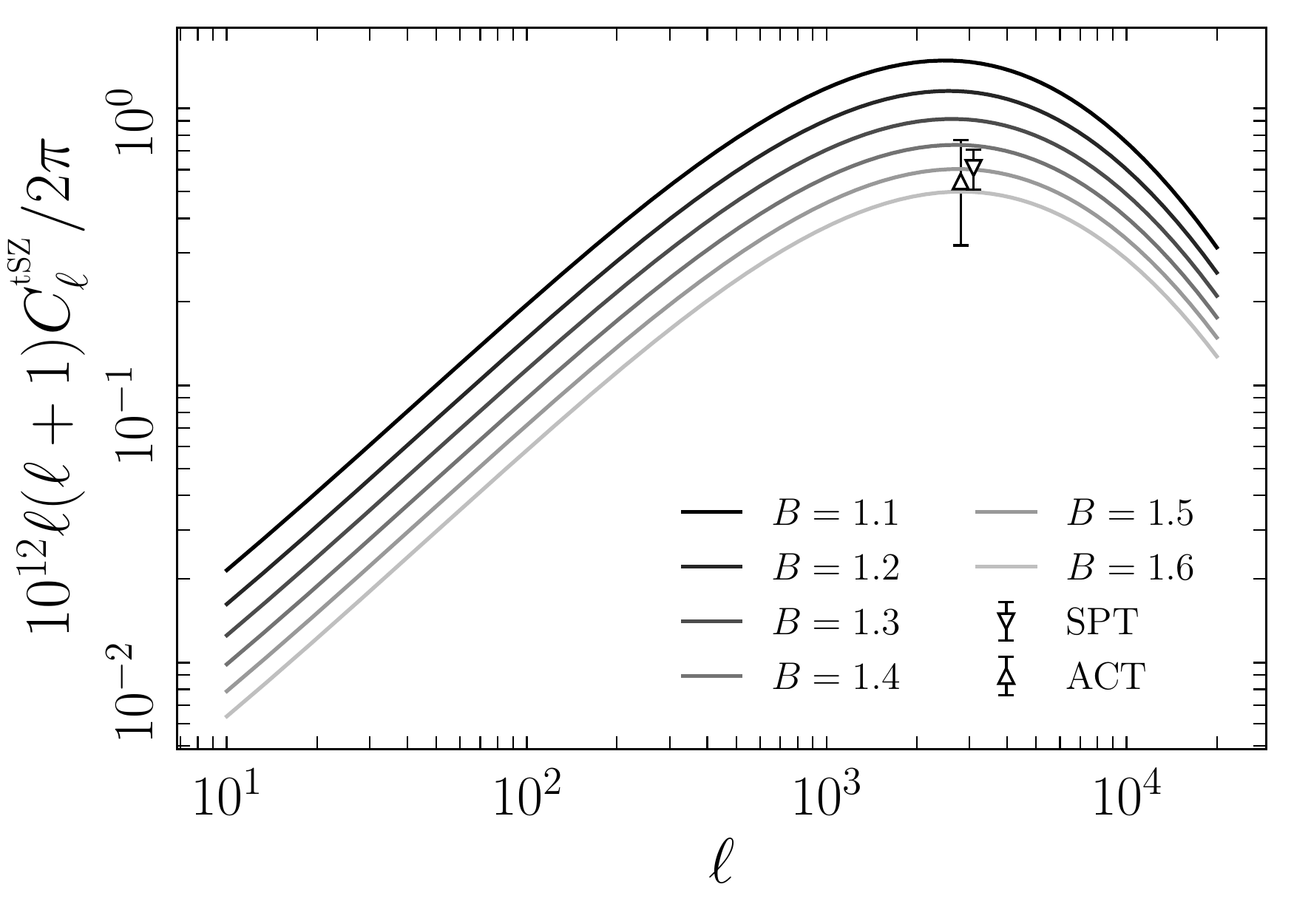} \includegraphics[width=5.7cm]{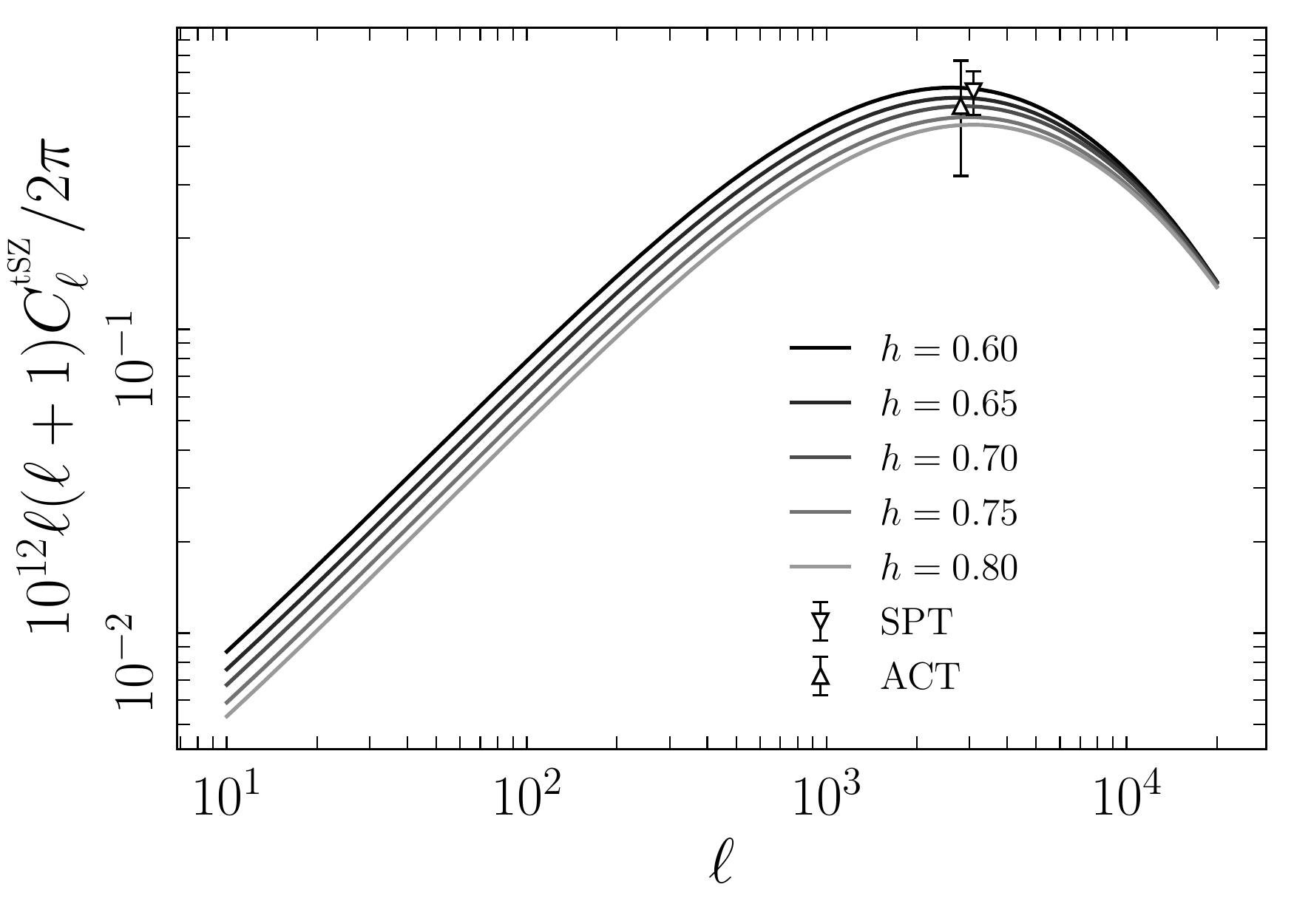} \includegraphics[width=5.7cm]{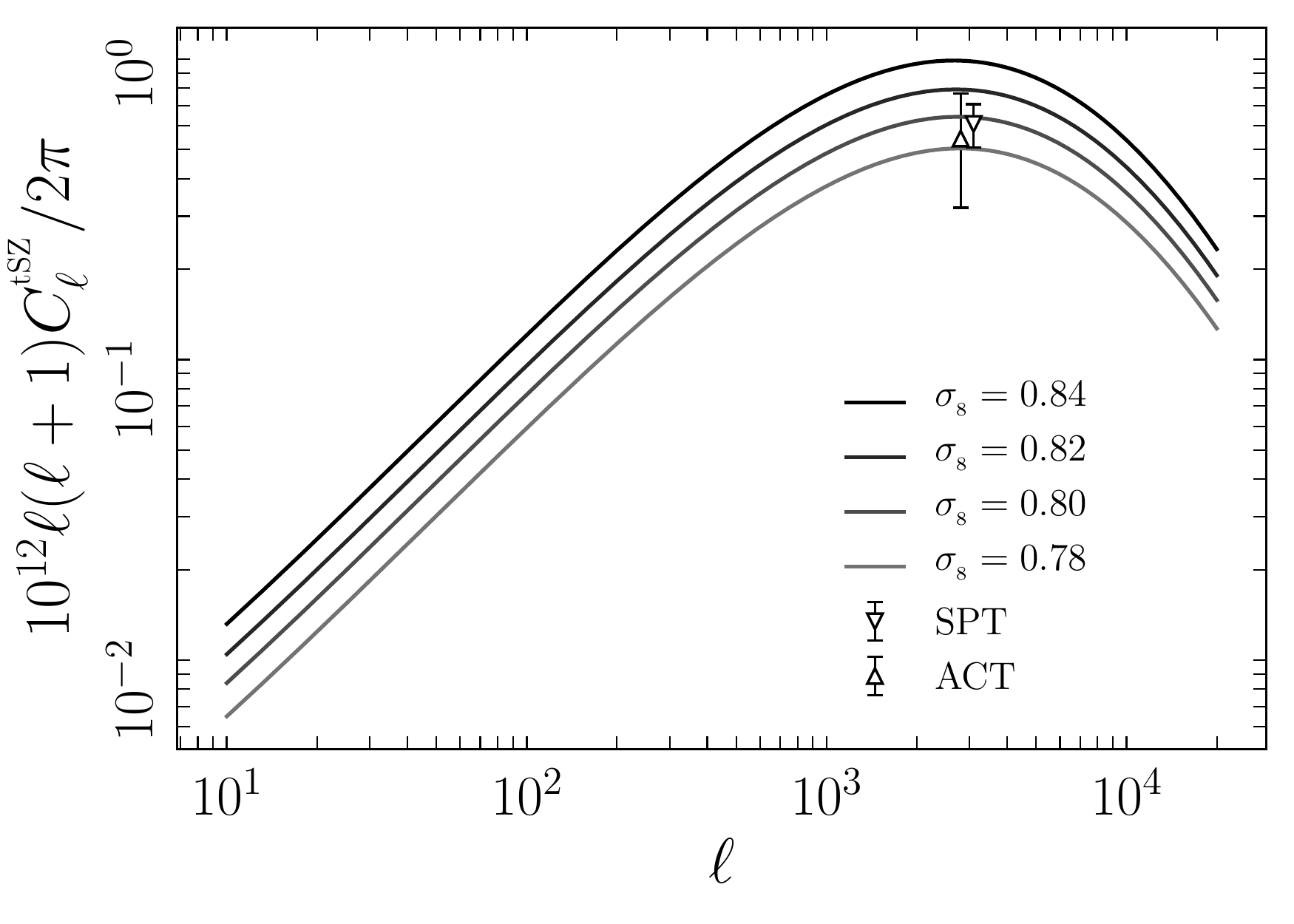}
 \includegraphics[width=5.7cm]{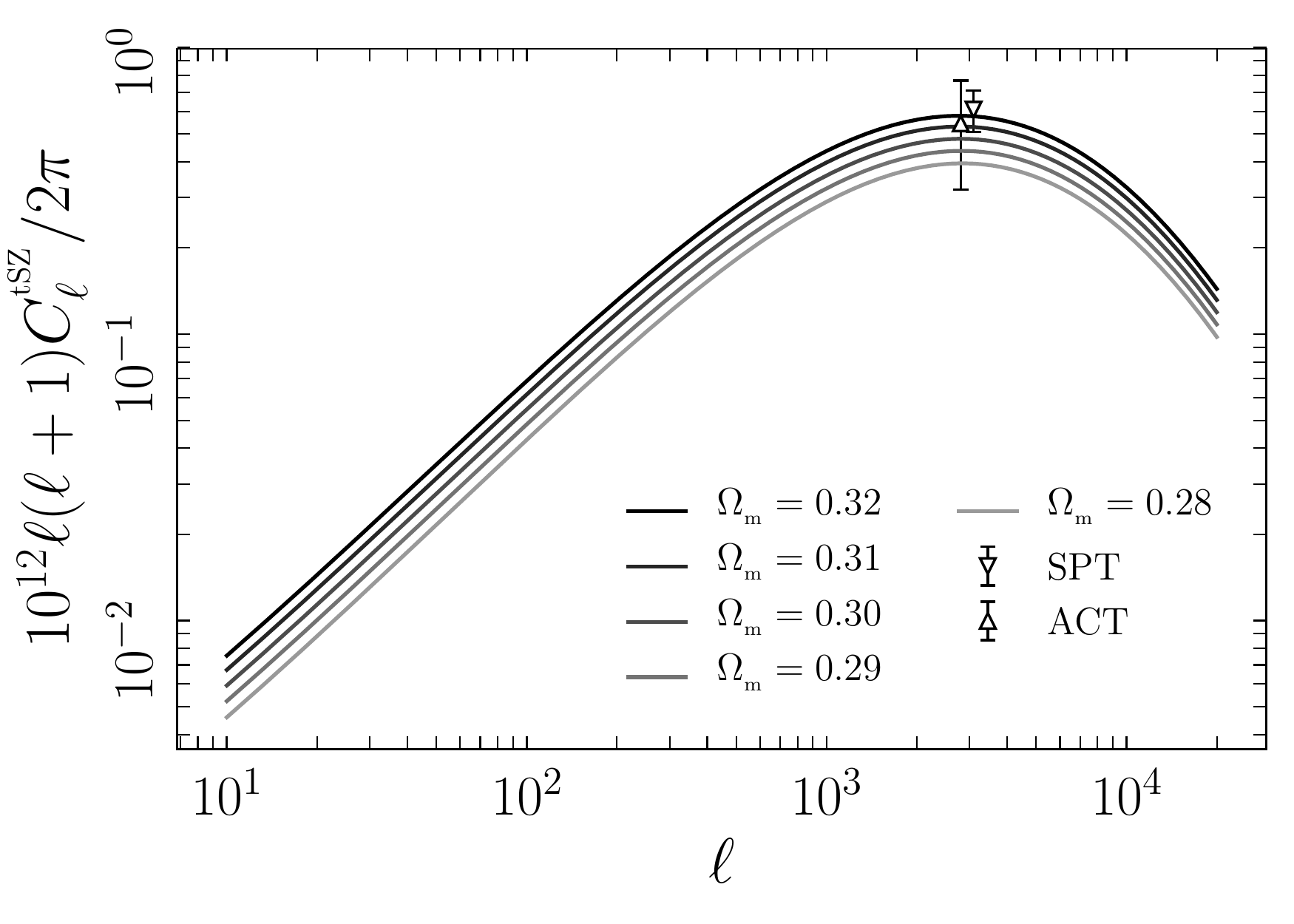}\includegraphics[width=5.7cm]{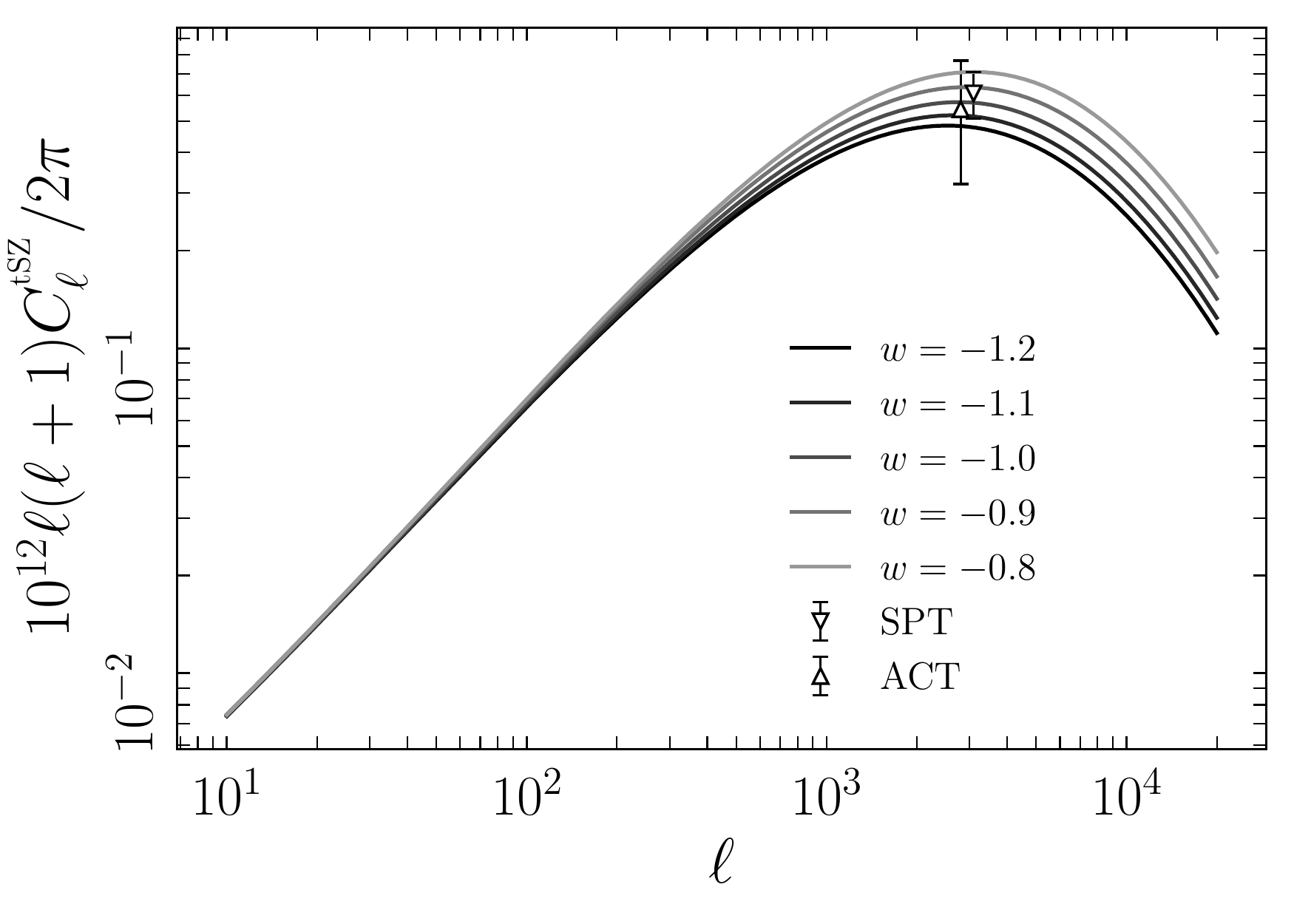} \includegraphics[width=5.7cm]{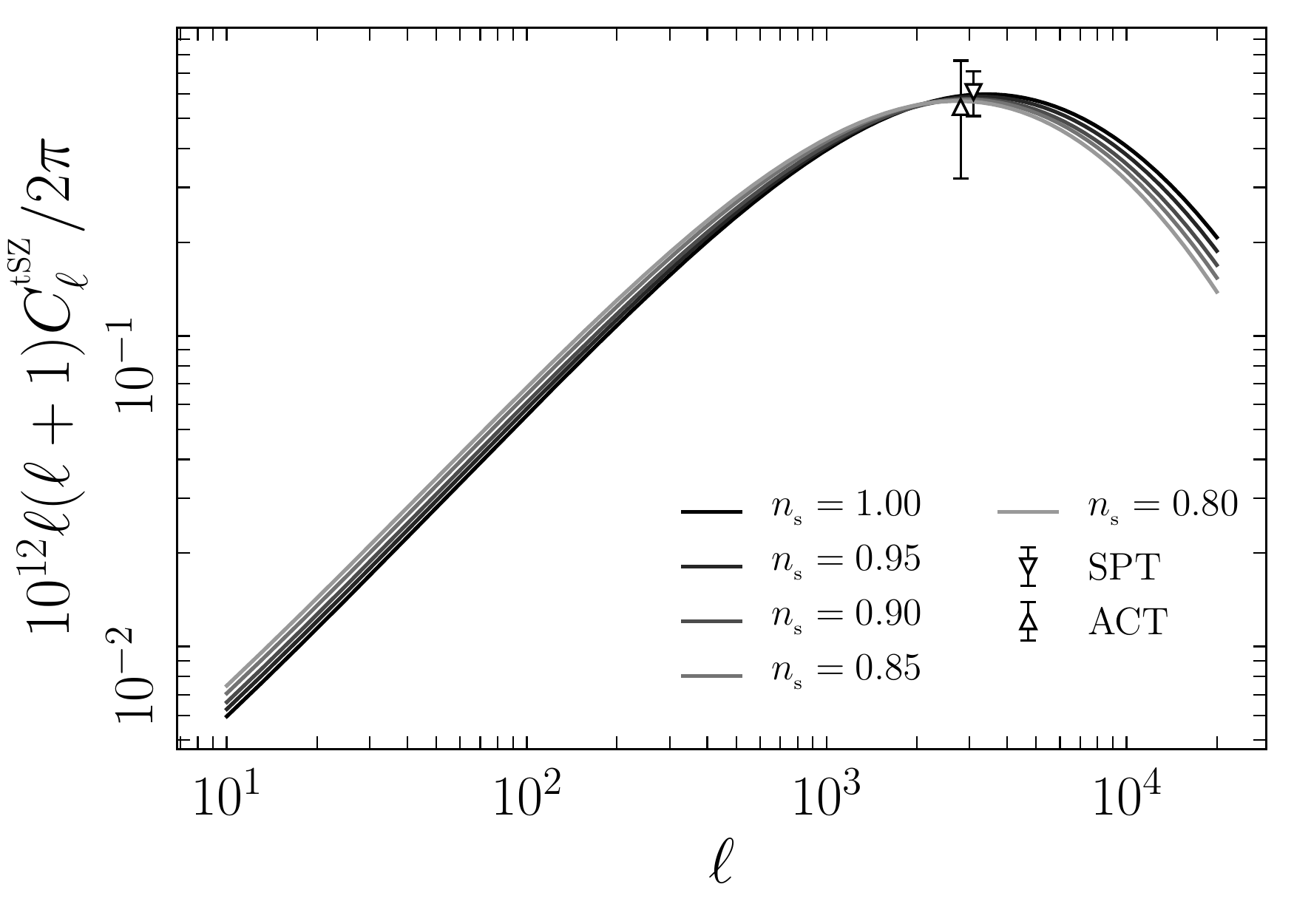}
 \caption{Dependence of the tSZ power spectrum on the mass bias $B$ as
 well as on the cosmological parameters $h$, $\sigma_8$,
 $\Omega_\mathrm{m}$, $w$, and $n_\mathrm{s}$. We hold $\sigma_8$ fixed by
adjusting the primordial scalar amplitude parameter $A_s$ in all but the
 top-left and top-right panels. The fiducial model is the same as in
 figure~\ref{fig:VaryingHMF}.\label{fig:Influence-of-various}}
\end{figure*}

In figure \ref{fig:Influence-of-various}, we show how the cosmological
parameters and the mass bias affect the tSZ power spectrum, computed with the \cite{Tinker:2008ff} HMF evaluated at $M_{\mathrm{500c}}$ so that no mass conversion is  needed. When we vary
$h$, $\Omega_\mathrm{m}$, $w$, and $n_s$, we hold $\sigma_8$ fixed by
adjusting the primordial scalar amplitude parameter $A_s$. For the
multipole range of interest ($\ell<10^{3}$), the EoS of dark energy $w$ (bottom
middle panel) and the spectral index $n_{{\mathrm{s}}}$ (bottom right
panel) have a minor effect on the amplitude of the tSZ power spectrum,
in agreement with \cite{Komatsu:2002wc}.
The mass bias $B$ as well as $h$, $\sigma_{8}$, and
$\Omega_{{\mathrm{m}}}$ affect the amplitude of the power spectrum
significantly. We find that the scaling of the power spectrum is well
approximated by
\begin{equation}
C_{\ell}^{{\mathrm{tSZ}}}\propto\sigma_{{{\mathrm{8}}}}^{8.1}\Omega_{{\mathrm{m}}}^{3.2}B^{-3.2}h^{-1.7}\,\,\,\,\mathrm{for}\,\,\,\,\,\mathrm{\ell}\lesssim10^{3}.\label{eq:scaling}
\end{equation}
The dependence on $\sigma_8$ and $\Omega_\mathrm{m}$ agrees with that of
\cite{Aghanim:2015eva}. We now add extra dependence on $B$ and $h$.
We can understand the dependence on $B$ approximately by looking at how
$y_\ell$ at small multipoles scales as $B$, i.e., $y_{\ell}\propto
r_{500}^3 M_{500c}^{2/3+0.12}$, which gives
$C_{\ell}^{{\mathrm{tSZ}}}\propto B^{-3.6}$. 
In this paper we simplify Eq.~\eqref{eq:scaling} as
\begin{equation}
F\equiv\sigma_{{{\mathrm{8}}}}\left(\Omega_{{\mathrm{m}}}/B\right)^{0.40}h^{-0.21},\label{eq:F}
\end{equation}
and find a constraint on $F$. 
\section{Maximum likelihood analysis}
\label{sec:MLA}
We derive cosmological constraints from the power spectrum of the
\cite{Aghanim:2015eva} Compton $y$ map. To this end we sample the
parameter space by the Markov Chain Monte Carlo (MCMC) method and
extract joint posterior probability distributions of the parameters. 
\begin{table*}
\begin{centering}
\setcellgapes{2pt}\makegapedcells
\begin{tabular}{ccccccccc}
\hline 
$\ell_{{\mathrm{eff}}}$ & $10^{12}\hat{D}_{\ell}^{y^{2}}$ & $\sigma_{\ell}^{y^{2}}$ & $10^{12}\hat{D}_{\ell}^{{\mathrm{RC}}}$  & $\sigma_{\ell}^{{\mathrm{RC}}}$ & $10^{12}\hat{D}_{\ell}^{{\mathrm{CIB}}}$ & $10^{12}\hat{D}_{\ell}^{{\mathrm{RS}}}$ & $10^{12}\hat{D}_{\ell}^{{\mathrm{IR}}}$ & $10^{12}\hat{D}_{\ell}^{{\mathrm{CN}}}$ \tabularnewline
\hline 
$10$  & 0.00508 & 0.00629  & 0.000421 & 0.000160 & 0.000000 & 0.000043 & 0.000007 & 0.000001\tabularnewline
$13.5$ & 0.00881 & 0.00615 & 0.000710 & 0.000192 & 0.000000 & 0.000142 & 0.000024 & 0.000001\tabularnewline
$18$ & 0.01363  & 0.00579  & 0.001251 & 0.000254 & 0.000000  & 0.000296  & 0.000048 & 0.000002\tabularnewline
$23.5$ & 0.02961 & 0.00805 & 0.002837 & 0.000446 & 0.000000  & 0.000400 & 0.000073 & 0.000004\tabularnewline
$30.5$ & 0.02241 & 0.00521 & 0.003933 & 0.000460 & 0.000902  & 0.000541 & 0.000111 & 0.000006\tabularnewline
$40$ & 0.02849 & 0.00464 & 0.005969 & 0.000510 & 0.002010 & 0.001056 & 0.000224 & 0.000010 \tabularnewline
52.5 & 0.04276 & 0.00468 & 0.010318 & 0.000672 & 0.003119 & 0.001647 & 0.000449 & 0.000018\tabularnewline
68.5 & 0.04580 & 0.00429 & 0.014045 & 0.000699 & 0.006278  & 0.002787 & 0.000837 & 0.000030\tabularnewline
89.5 & 0.07104 & 0.00454 & 0.024061 & 0.000896 & 0.012242 & 0.004306 & 0.001400 & 0.000052\tabularnewline
117 & 0.11914 & 0.00562  & 0.032976 & 0.000936 & 0.021584 & 0.006842 & 0.002701 & 0.000089\tabularnewline
152.5 & 0.15150 & 0.00594 & 0.047100 & 0.001020 & 0.045915 & 0.011264  & 0.004721 & 0.000153\tabularnewline
198 & 0.19390 & 0.00611 & 0.062380 & 0.001040 & 0.070582 & 0.016744 & 0.008115  & 0.000262\tabularnewline
257.5 & 0.28175 & 0.00687 & 0.081730 & 0.001030 & 0.119786 & 0.027345 & 0.014618 & 0.000456\tabularnewline
335.5 & 0.39837 & 0.00824 & 0.101911 & 0.000978 & 0.211686 & 0.043275 & 0.024893 & 0.000815\tabularnewline
436.5 & 0.56743 & 0.00958 & 0.117412 & 0.000860 & 0.332863 & 0.070587 & 0.051570 & 0.001503\tabularnewline
567.5 & 0.76866 & 0.01242 & 0.132234 & 0.000769 & 0.434931 & 0.115356 & 0.107293 & 0.002934\tabularnewline
738 & 1.11010 & 0.01650 & 0.143214 & 0.000642 & 0.602030 & 0.154926 & 0.197053 & 0.006334\tabularnewline
959.5 & 1.66140 & 0.02400 & 0.156202 & 0.000544 & 0.754733 & 0.207200 & 0.361713 & 0.016171\tabularnewline
\hline 
1247.5 & 2.52170 & 0.04170 & 0.175341 & 0.000492 & 1.029014 & 0.287652 & 0.681036 & 0.054883\tabularnewline
1622 & 4.58510 & 0.09870 & 0.283969 & 0.000900 & 1.357567 & 0.410274 & 1.295272 & 0.301480\tabularnewline
2109 & 12.2690 & 0.40100 & 1.363680 & 0.003650 & 1.850146 & 0.657659 & 2.534448 & 3.738250\tabularnewline
\hline 
2742 & 165.600 & 23.6000 & 54.69000 & 2.310000 & 2.629002 & 1.117189 & 4.545315 & 183.2673\tabularnewline
\hline 
\end{tabular}
\par\end{centering}
\caption[]{Planck 2015 data points and Gaussian error bars of the $y^{2}$ power
 spectrum $\hat{C}_{\ell}^{y^{2}}$, the tSZ power
 spectrum from resolved  clusters $\hat{C}_{\ell}^{\mathrm{RC}}$, and templates for the
 foreground contributions: CIB, IR, and RS. The last column is the power
 spectrum of correlated noise (CN). These data were taken from
 \cite{Aghanim:2015eva}. Note that we used $D_{\ell}\equiv [\ell(\ell+1)/2\pi] C_{\ell}$.}  \label{tab:Data-used-in}
\end{table*}

We use the effective multipole range of
$10\leq\ell_{{\mathrm{eff}}}\leq959.5$. The effective multipoles are
defined to be the middle of the eighteen bins spanning this interval and
used in the \cite{Ade:2013qta} and \cite{Aghanim:2015eva} (PLC15) tSZ
analyses.

We assume a flat universe and the standard number of neutrino species
$N_{{\mathrm{eff}}}=3.046$, with $\Sigma m_{\nu}=0.06\,\mathrm{eV}$. Although neutrinos play an important role on non-linear scales because they slow down gravitational collapse, we ignore their effect on the halo mass function in this paper (as was done in the Planck SZ analyses). Given the low value of the neutrino mass we consider, this is an acceptable approximation: a careful treatment would affect the determination of $F$ by less than one percent, see \cite{Costanzi:2013bha}. 

We vary $B$, $A_s$, $n_s$, and $w$, as well
as the optical depth to
electron scattering during re-ionisation $\tau_{{\mathrm{reio}}}$,
the angular size of the sound horizon at decoupling $\theta_{s}$,
the baryon density $\Omega_{{\mathrm{b}}}h^{2}$, 
the cold dark matter density
$\Omega_{{\mathrm{c}}}h^{2}$, and the nuisance parameters
($A_\mathrm{CIB}$, $A_\mathrm{IR}$, $A_\mathrm{RS}$) that will be
described below. The other parameters, $h$, $\sigma_{8}$, and
and $\Omega_{{\mathrm{m}}}$, are derived parameters.

We split these parameters into the so-called ``fast'' and ``slow'' parameters as
\begin{equation}
\underbrace{A_{\mathrm{s}},n_{\mathrm{s}},\tau_{{\mathrm{reio}}},\theta_{s},\Omega_{{\mathrm{b}}}h^{2},\Omega_{{\mathrm{c}}}h^{2},w,B}_{C_{\ell}^{\mathrm{tSZ}}:\,\,slow\,\,\mathrm{param.}},\underbrace{A_{{\mathrm{CIB}}},A_{{\mathrm{IR}}},A_{{\mathrm{RS}}}}_{C_{\ell}^{\mathrm{FG}}:\,\,fast\,\,\mathrm{param.}}\,.
\end{equation}

This splitting makes MCMC exploration of the parameter space
 efficient \citep{Lewis:2013hha}. We use \verb|Montepython|
\citep{2013JCAP...02..001A} for the sampling. We impose weak uniform priors on the parameters to avoid unrealistic
regions of the parameter space. See table
\ref{tab:Uniform-priors-imposed} for the priors we adopt. The prior on the mass bias $B$ is motivated by the scatter of the results from numerical simulations (it corresponds to $0.1<b<0.4$). For the other
parameters, the allowed range is wide enough so that changing the upper
or lower bound does not affect our posterior likelihood. 

We start by fitting the {\it total} power spectrum of the Compton $y$
map, $\hat{C}_{\ell}^{y^{2}}$, which includes contributions from tSZ, three
foreground components (the cosmic infrared background (CIB), radio
sources (RS) and infrared point sources (IR)), and a correlated noise
(CN) term. Our model is 
\begin{align}
C_{\ell}^{y^{2}} & =C_{\ell}^{{\mathrm{tSZ}}}+A_{{\mathrm{CIB}}}\hat{C}_{\ell}^{{\mathrm{CIB}}}+A_{{\mathrm{IR}}}\hat{C}_{\ell}^{{\mathrm{IR}}}+A_{{\mathrm{RS}}}\hat{C}_{\ell}^{{\mathrm{RS}}}+A_{{\mathrm{CN}}}\hat{C}_{\ell}^{{\mathrm{CN}}},\label{eq:model-2}
\end{align}
where $\hat{C}_{\ell}^{{\mathrm{CIB}}}$, $\hat{C}_{\ell}^{{\mathrm{IR}}}$,
$\hat{C}_{\ell}^{{\mathrm{RS}}}$, and $\hat{C}_{\ell}^{{\mathrm{CN}}}$ are
templates of the foreground and correlated noise terms, respectively (see table
\ref{tab:Data-used-in}).
Since the correlated noise term dominates over the other terms at high multipoles, we
use the highest multipole data at $\ell=2742$ to determine
$A_{\mathrm{CN}}$, i.e., $A_{{\mathrm{CN}}}=\hat{C}_{2742}^{y^{2}}/\hat{C}_{2742}^{{\mathrm{CN}}}=0.903$.
We then fit for and marginalise over
the amplitudes of the foreground term, $A_{{\mathrm{CIB}}},A_{{\mathrm{IR}}},A_{{\mathrm{RS}}}$.

In fact, there is a physical upper bound on the total foreground power;
namely, the sum of the tSZ power from {\it resolved}
sources\footnote{The data for $\hat{C}_{\ell}^{{\mathrm{RC}}}$ and the
error bars $\sigma_{\ell}^{{\mathrm{RC}}}$ are also reported in table
\ref{tab:Data-used-in} \citep[see][for details]{Ade:2013qta,Aghanim:2015eva}.}
$\hat{C}_\ell^{\rm RC}$ and the
total foreground power cannot exceed the $\hat{C}_\ell^{y^2}$ data. Thus,
\begin{equation}
A_{{\mathrm{CIB}}}\hat{C}_{\ell}^{{\mathrm{CIB}}}+A_{{\mathrm{IR}}}\hat{C}_{\ell}^{{\mathrm{IR}}}+A_{{\mathrm{RS}}}\hat{C}_{\ell}^{{\mathrm{RS}}}+A_{{\mathrm{CN}}}\hat{C}_{\ell}^{{\mathrm{CN}}}<\hat{C}_{\ell}^{{y}^{2}}-\hat{C}_{\ell}^{\mathrm{{RC}}}.\label{eq:rejec}
\end{equation}
We use this upper bound as follows. At each step of the MCMC, we ensure
that the foreground amplitudes $\left\{
A_{{\mathrm{CIB}}},A_{{\mathrm{IR}}},A_{{\mathrm{RS}}}\right\} $ satisfy
Eq.~\eqref{eq:rejec} in the seven
multipole bins between $\ell_{{\mathrm{eff}}}=257.5$
and $1247.5$; the proposed step in the MCMC is
rejected otherwise. This multipole range is chosen because above $\ell_{{\mathrm{eff}}}=1247.5$ $\hat{C}_{\ell}^{{\mathrm{RC}}}$
is significantly affected by the resolution of the y-map, while bellow $\ell_{{\mathrm{eff}}}=257.5$, statistical
and systematic uncertainties are important and Eq.~\eqref{eq:rejec} is
no longer applicable.\\
\\

The likelihood is computed according
to $-2\ln\mathcal{L}=\chi^{2}+\ln|M|+{\rm const.}$ with 
\begin{equation}
\chi^{2}\equiv\sum_{a\leq a^{\prime}}\left(C_{\ell_{{\mathrm{eff}}}^{a}}^{y^{2}}-\hat{C}_{\ell_{{\mathrm{eff}}}^{a}}^{y^{2}}\right)\left[M^{-1}\right]_{aa^{\prime}}\left(C_{\ell_{{\mathrm{eff}}}^{a^{\prime}}}^{y^{2}}-\hat{C}_{\ell_{{\mathrm{eff}}}^{a^{\prime}}}^{y^{2}}\right),\label{eq:likelihood}
\end{equation}
where $a,a^{\prime}$ are indices for the multipole bins running from
$a=1$ to $a=18$, $C_{\ell_{{\mathrm{eff}}}^{a}}^{y^{2}}$
is given by Eq.~\eqref{eq:model-2},
$\hat{C}_{\ell_{{\mathrm{eff}}}^{a}}^{y^{2}}$ are the data points,
$M$ is the binned covariance matrix, and $|M|$ is its determinant. The elements of the binned covariance matrix are
\begin{equation}
M_{aa^{\prime}}=\left(\sigma_{\ell_{{\mathrm{eff}}}^{a}}^{y^{2}}\right)^{2}\delta_{aa^{\prime}}+\frac{\ell_{{\mathrm{eff}}}^{a}\left(\ell_{{\mathrm{eff}}}^{a}+1\right)\ell_{{\mathrm{eff}}}^{a^{\prime}}(\ell_{{\mathrm{eff}}}^{a^{\prime}}+1)}{4\pi^{2}}\frac{T_{aa^{\prime}}}{4\pi f_{{\mathrm{sky}}}},\label{eq:covmat}
\end{equation}
where $\sigma_{\ell_{{\mathrm{eff}}}^{a}}^{y^{2}}$
are the measured error bars (third column of table \ref{tab:Data-used-in}) which comes from both the sampling variance  and Gaussian instrumental noise,
$f_{{\mathrm{sky}}}=0.47$ is the Planck sky coverage,
and $T_{aa^{\prime}}$ is the binned trispectrum given by
\begin{equation}
T_{aa^{\prime}}=\sum_{\ell\in a}\sum_{\ell^{\prime}\in a^{\prime}}\frac{T_{\ell\ell^{\prime}}}{N_{a}N_{a^{\prime}}},\label{eq:Taa-1}
\end{equation}
where $a$ and $a^{\prime}$ denote two multipole bins containing
respectively $N_{a}$ and $N_{a^{\prime}}$ multipoles. The unbinned trispectrum $T_{\ell\ell^{\prime}}$ 
is assumed to be dominated by the tSZ effect contribution. We follow
\cite{Komatsu:2002wc} and calculate
\begin{equation}
T_{\ell\ell^{\prime}}=\int\mathrm{d}z\frac{dV}{dzd\Omega}\int\mathrm{d}\ln M\frac{dn}{d\ln M}\left|y_{\ell}\left(M,z\right)\right|^{2}\left|y_{\ell^{\prime}}\left(M,z\right)\right|^{2},\label{eq:tll-1}
\end{equation}
where the redshift and mass ranges are the same as for the power spectrum, see Eq.~\eqref{eq:cltSZ-1}. 

\begin{figure*}
\begin{centering}
\includegraphics[width=7cm]{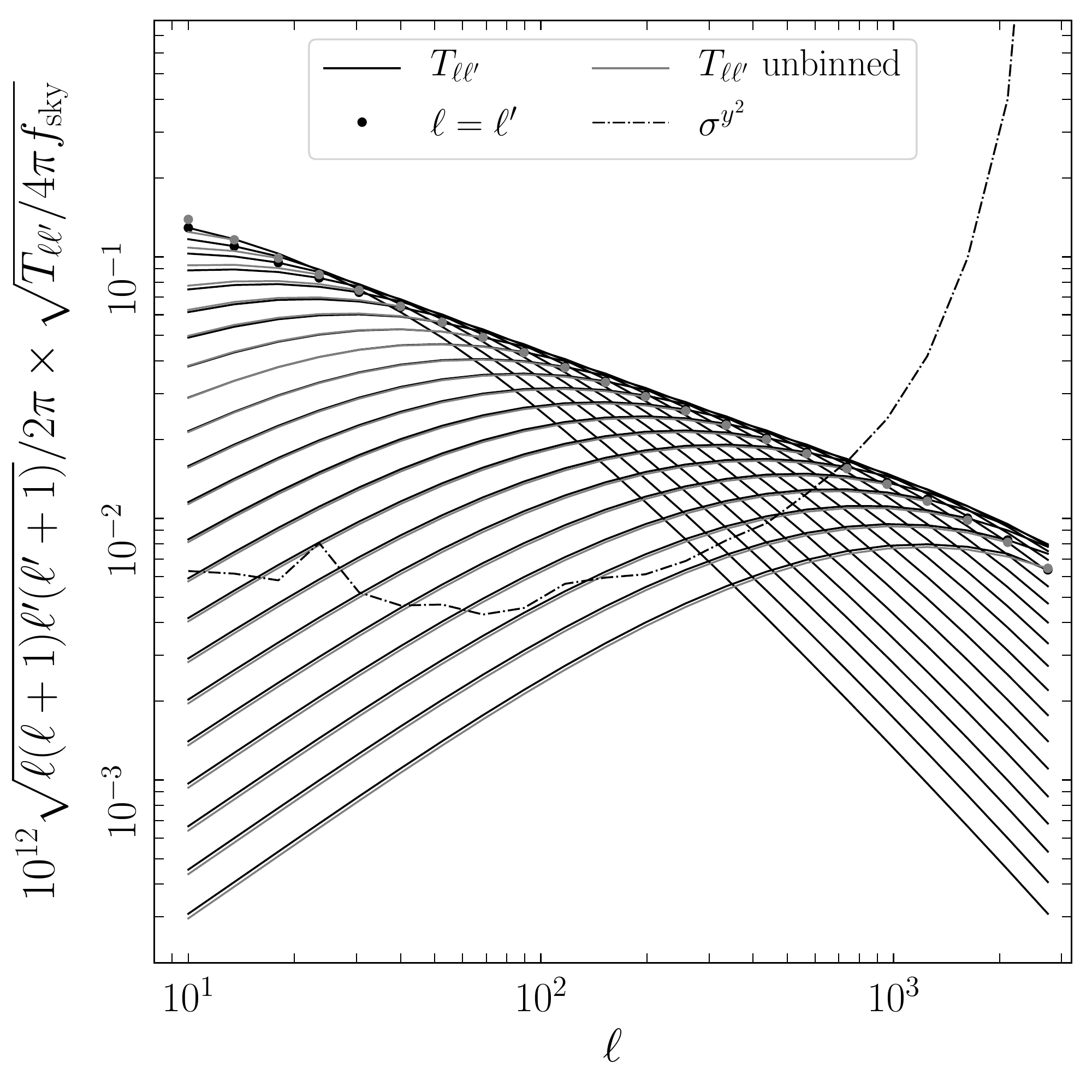}
\includegraphics[width=7cm]{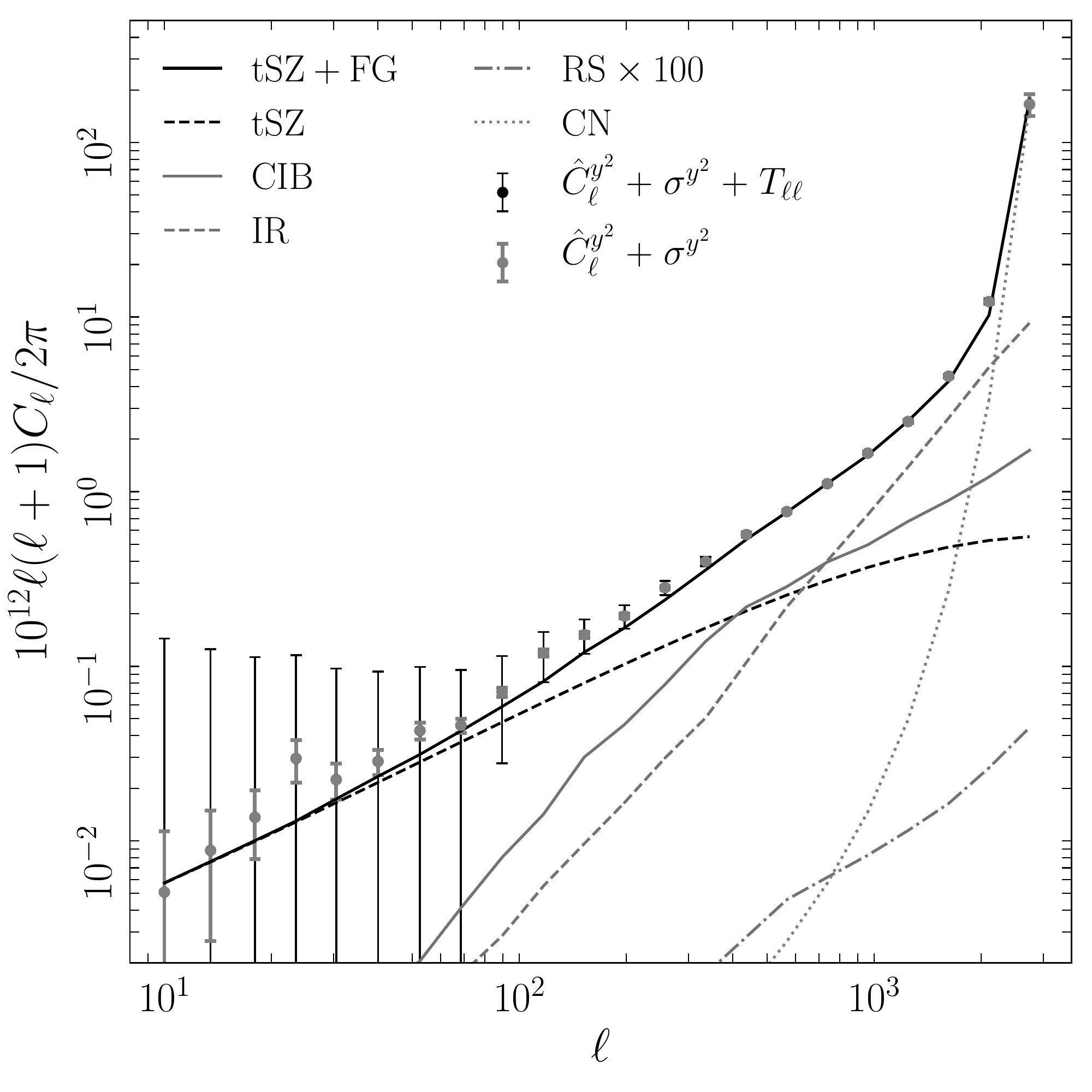}
\par\end{centering}
 \caption{(Left panel) tSZ power spectrum covariance matrix, for the multipole values $\ell_{\mathrm{eff}}$ listed in table \ref{tab:tSZBF}. The
 dot-dashed line shows the Gaussian term $\sigma_\ell^{y^2}$, while the thick and thin solid
 lines show the binned and unbinned trispectrum terms
 $10^{12}\sqrt{\ell(\ell+1)\ell'(\ell'+1)/4\pi^2}\times
 \sqrt{T_{\ell\ell'}/4\pi f_{\rm sky}}$, respectively (see
 Eq.\eqref{eq:covmat}). The filled
 circles show the diagonals of the trispectrum, and the lines show
 off-diagonal terms. (Right panel) The best-fitting tSZ (black dashed
 line), CIB (grey solid), IR (grey dashed), RS (times 100; dot-dashed),
 and correlated noise (CN; dotted) power spectra. The black solid line shows the sum of
 all the components. The filled circles show the data points used in this
 work, with the Gaussian and total diagonal elements of the covariance
 matrix shown in the grey and black error bars, respectively. The
 parameters of the best-fitting model are: $\sigma_8=0.78$,
 $\Omega_{\mathrm{m}}=0.31$, $h=0.69$,  $B=1.41$ (i.e., $F=0.460$),
 $n_{\mathrm{s}}=0.96$, $A_\mathrm{CIB}=0.66$, $A_\mathrm{IR}=2.04$, and
 $A_\mathrm{RS}=0.0004$.}
\label{fig:Reference-trispectrum-for-1}
\end{figure*}

In the left panel of figure \ref{fig:Reference-trispectrum-for-1} we compare the
trispectrum and Gaussian contributions to the covariance matrix.
As found by \cite{Komatsu:2002wc}, the trispectrum dominates over the
Gaussian term at low multipoles (when binned linearly in
multipoles). This is due to large massive halos at low redshift covering
a large fraction of the sky. Adding a few large halos in the sky
increase the tSZ power significantly at many multipoles, producing a
large multipole-to-multipole correlation. 
We find that the binned and unbinned trispectra are similar; thus, to
save computational time we use the unbinned trispectra evaluated
at $\ell_{\rm eff}$.

As both the power spectrum and trispectrum depend on the cosmological
parameters, we compute them at each step of the MCMC sampling.
In principle, we would also need to vary the Gaussian term of the covariance matrix at each step; however, since the total variance is dominated by the
trispectrum at low multipoles and by Gaussian instrumental noise at high multipoles, we use the values derived
by the PLC15 analysis.
\section{Results}\label{sec:Results}
In subsection \ref{subsec:Revisiting} we compare constraints on $F$ with
and without trispectrum in the covariance matrix. In subsection
\ref{subsec:mass_bias} we obtain the mass bias that reconciles the tSZ and
primary CMB data within the context of $\Lambda$CDM. Finally, in
subsection \ref{subsec:DE} we present our results on dark energy.
\subsection{Constraints on $F$}
\label{subsec:Revisiting}
To illustrate importance of the trispectrum, we first obtain a
constraint on $F$ without it. Marginalising over all the relevant
cosmological, foreground, and mass bias parameters, we find
$F=0.473\pm0.005$ (68\%~CL). 
\begin{figure*}
\begin{centering}
\includegraphics[width=13.5cm]{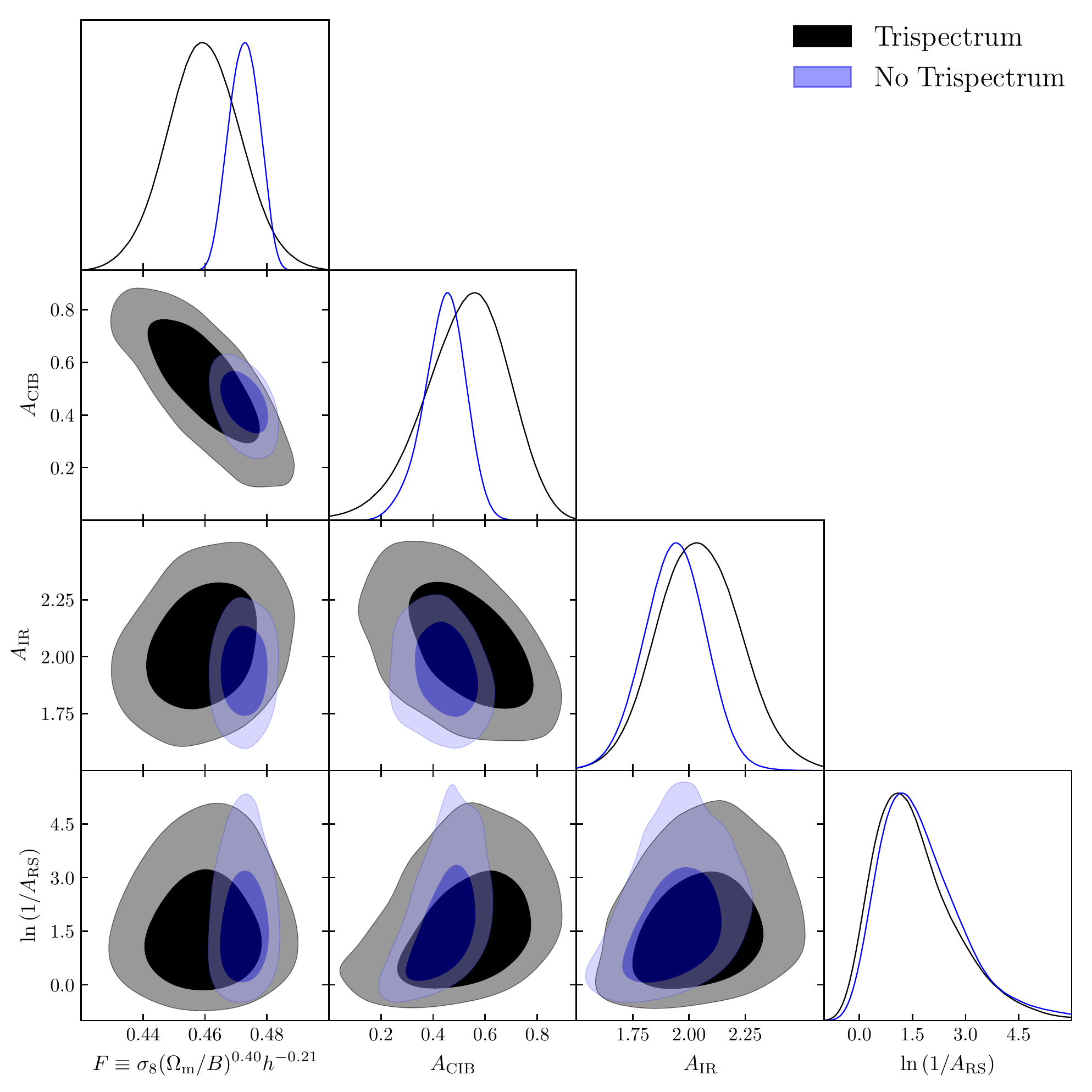}
\par\end{centering}
\caption{Marginalised (1D and 2D) joint posterior probability
 distributions of the $F$-parameter (Eq.~\eqref{eq:F}) and the
 foreground amplitudes (CIB, IR, and RS). The black and blue contours
 show the results with and without trispectrum in the covariance,
 respectively.  \label{fig:revisiting}}
\end{figure*}
\begin{figure*}
\begin{centering}
\includegraphics[width=8cm]{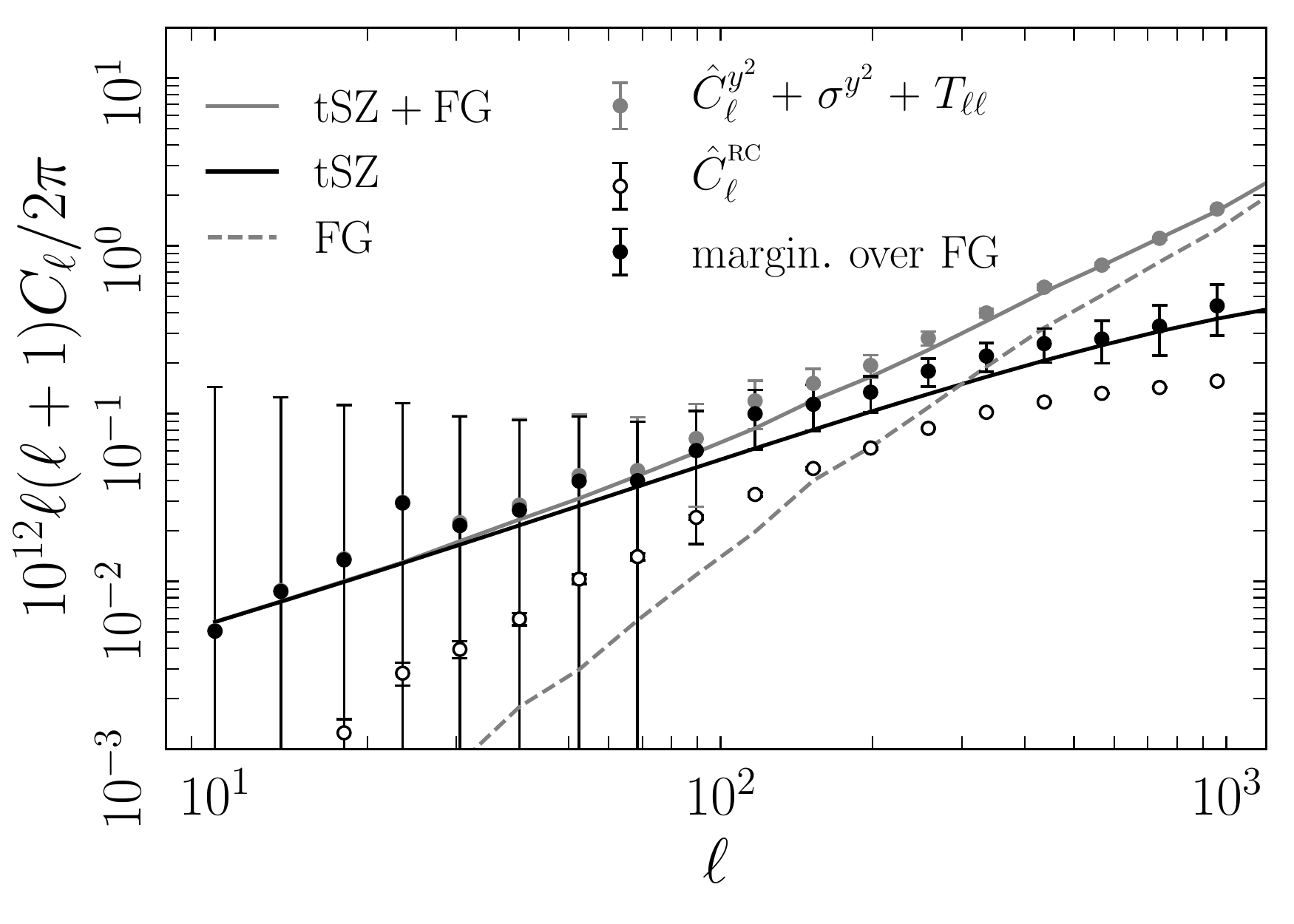}\includegraphics[width=8cm]{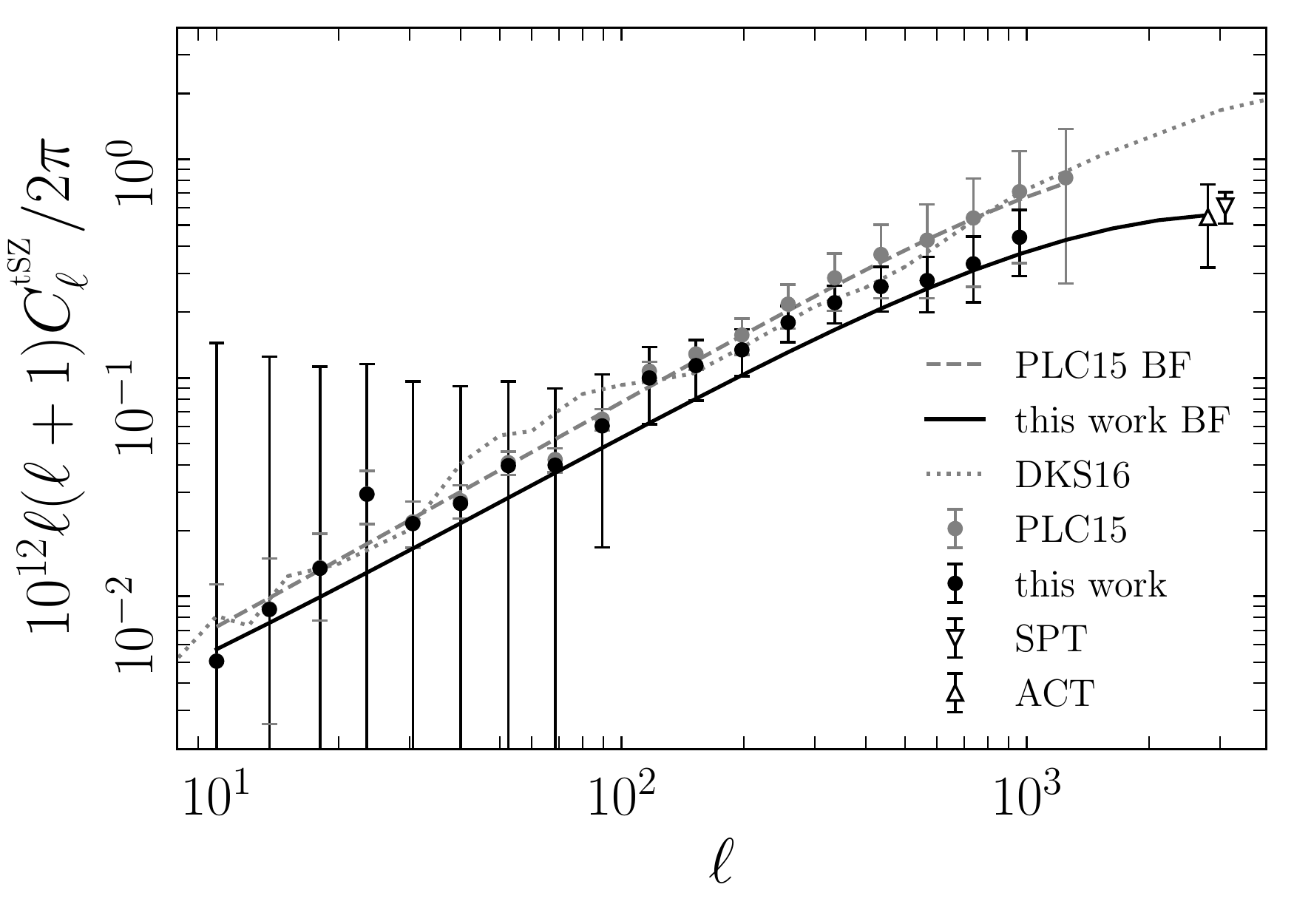}
\par\end{centering}
\caption{(Left panel) Best-fitting tSZ (solid black line) and total
 foreground (dashed line) power spectra. The solid grey line shows their
 sum. The open circles show the contribution from resolved clusters. The
 grey circles show the original data points with error bars including
 the trispectrum, whereas the black filled circles show the data points
 with the foreground spectra marginalised over. (Right panel) The black
 line and circles are the same as those in the left panel. See figure \ref{fig:Reference-trispectrum-for-1} for the
 best-fitting model parameters, and table \ref{tab:tSZBF} for the
 tabulated values. The grey
circles show the foreground-marginalised data points obtained by the
 Planck 2015 SZ analysis. We also show the ACT and SPT data points for
 comparison, as well as the simulation results of \protect\cite{Dolag:2015dta} (DKS16, grey dotted line). \label{fig:CodeComp}}
\end{figure*}
When the trispectrum is included, the 68\% C.L. error bar increases by a
factor of more than two. We find 
 $F=0.460\pm0.012~(68\%~{\rm CL})$,
and $A_{{\mathrm{CIB}}}=0.53\pm0.16$, $A_{{\mathrm{IR}}}=2.05\pm0.18$, 
$A_{{\mathrm{RS}}}=0.34_{-0.34}^{+0.07}$.

In the right panel of figure \ref{fig:Reference-trispectrum-for-1} we show the
best-fitting power spectra of the tSZ and foregrounds, as well
as the original $\hat{C}_\ell^{y^{2}}$ data points.  We show the correlations among $F$ and
foreground parameters in figure \ref{fig:revisiting}. In the left panel of figure~\ref{fig:CodeComp} we show the derived tSZ
power spectrum data points (with the foregrounds subtracted and
marginalised over; black points, see table \ref{tab:tSZBF} for the data). In the right panel of figure~\ref{fig:CodeComp} we compare our
derived tSZ power with those of the PLC15 (grey points) and the ACT
\citep{Hasselfield:2013wf} and SPT \citep{George:2014oba} measurements
at $\ell=3000$. Our tSZ data points at $\ell\gtrsim 300$ are
significantly smaller than the PLC15, and we attribute this difference
to our including the trispectrum in the analysis.  We also show the tSZ power spectrum computed from the \verb|Magneticum|
simulation \citep{Dolag:2015dta} with
$\Omega_\mathrm{m}=0.308$, $\sigma_8=0.815$, $h=0.704$, $n_s=0.963$, and
$\Omega_\mathrm{b}=0.0456$. While this power spectrum agrees well with the
PLC15 data points, as claimed by the authors, our best-fitting model is
approximately 2/3 of their power spectrum.

\begin{center}
\begin{table}
\begin{centering}
\setcellgapes{2pt}\makegapedcells
\begin{tabular}{cccc}
\hline 
$\ell_{{\mathrm{eff}}}$ & $10^{12}D_{\ell}^{{\mathrm{tSZ}}}$& $\sigma_{\ell}^{\mathrm{tSZ}}$&$10^{12}D_{\ell}^{{\mathrm{tSZ}}}$ (BF)\tabularnewline
\hline 
$10$  &5.0496e-03& 1.3919e-01& 5.7234e-03\tabularnewline
$13.5$ & 8.7080e-03& 1.1643e-01& 7.5606e-03 \tabularnewline
$18$ &  1.3430e-02& 9.9155e-02& 9.9280e-03\tabularnewline
$23.5$ &  2.9325e-02& 8.6276e-02& 1.2834e-02\tabularnewline
$30.5$ & 2.1520e-02& 7.4827e-02& 1.6544e-02\tabularnewline
$40$ &  2.6605e-02& 6.4972e-02& 2.1591e-02 \tabularnewline
52.5 &  3.9626e-02& 5.6584e-02& 2.8250e-02\tabularnewline
68.5 &  3.9820e-02& 4.9513e-02& 3.6721e-02\tabularnewline
89.5 &  6.0241e-02& 4.3516e-02& 4.7775e-02\tabularnewline
117 & 9.9878e-02& 3.8669e-02& 6.2081e-02\tabularnewline
152.5 &1.1375e-01& 3.4993e-02& 8.0185e-02\tabularnewline
198 &  1.3429e-01& 3.2630e-02& 1.0271e-01\tabularnewline
257.5 &  1.7920e-01& 3.3977e-02& 1.3093e-01\tabularnewline
335.5 & 2.2076e-01& 4.3272e-02& 1.6575e-01\tabularnewline
436.5 &  2.6166e-01& 6.0483e-02& 2.0728e-01\tabularnewline
567.5 &2.7879e-01& 7.9377e-02& 2.5553e-01\tabularnewline
738 &  3.3226e-01& 1.1034e-01& 3.0982e-01\tabularnewline
959.5 &  4.3979e-01& 1.4774e-01& 3.6810e-01\tabularnewline
1247.5 & --& --& 4.2699e-01\tabularnewline
1622 & --& --& 4.8154e-01\tabularnewline
2109 & --& --&5.2553e-01\tabularnewline
2742 & --& --& 5.5249e-01\tabularnewline
\hline 
\end{tabular}
\par\end{centering}
\caption{tSZ power spectrum and total uncertainties with the foreground and correlated noise
 power spectra marginalised over. The last column is the best-fitting
 tSZ power spectrum (see right panel of figure \ref{fig:CodeComp}). Note that we used $D_{\ell}\equiv [\ell(\ell+1)/2\pi] C_{\ell}$. \label{tab:tSZBF}}
\end{table}
\end{center}
\subsection{Mass bias in $\Lambda$CDM}
\label{subsec:mass_bias}
Given a cosmological model, we can constrain the mass bias $B$ by
combining the tSZ likelihood with the Planck CMB data. Assuming
a flat $\Lambda$CDM and using the Planck CMB ``TT+lowP'' chains
\citep{Ade:2015xua}, we find
$\sigma_{{{\mathrm{8}}}}\Omega_{{\mathrm{m}}}^{0.40}h^{-0.21}=0.568\pm0.015$
(68\%~CL). Comparing this to the above constraint on $F$, we derive the
mass bias as 
$B=1.71\pm0.17~(68\%~{\rm CL})$,
or $(1-b)=0.58\pm0.06$.

The origin of this mass bias is not known. It is possible that a part of
the bias comes from non-thermal pressure support because $\tilde
M_{500\rm c}$ in the universal pressure profile was derived assuming
hydrostatic equilibrium with thermal gas pressure
\citep{2010A&A...517A..92A}. However, the value of $B$ we find appears
to be significantly larger than expected from both analytical calculations and some numerical
simulations \citep[e.g.,][and references
therein]{Shi:2014msa,Shi:2014lua}, but see \cite{Henson:2016eip} for the
simulation work pointing towards a larger mass bias. It is also possible that hydrostatic
equilibrium does not hold even when we add both thermal and non-thermal
pressure, though this is not expected to be significant for $M_{500\rm c}$
\citep{Lau:2009qm}. Finally, the bias may be due to non-physical effects
such as instrumental and analysis systematics of the X-ray data used to
derive $\tilde M_{500\rm c}$ in the universal pressure profile.

Observationally, the mass bias we find is consistent with a recent
weak lensing constraint, $(1-b)=0.73\pm0.10$
\citep{Penna-Lima:2016tvo}, as well as with the Planck tSZ cluster
number counts  analysis which yields $(1-b)=0.58\pm0.04$ \citep{Ade:2015fva}.

Alternatively, such a large mass bias required within the
context of $\Lambda$CDM hints that the mass bias is not a
culprit but a modification to $\Lambda$CDM, such as dark energy that is
different from a cosmological constant, may be needed.
\subsection[Constraints on $w_{{\mathrm{de}}}$ with the SZ
data]{Constraints on Dark Energy}
\label{subsec:DE}
\begin{figure*}
\begin{centering}
\includegraphics[width=14cm]{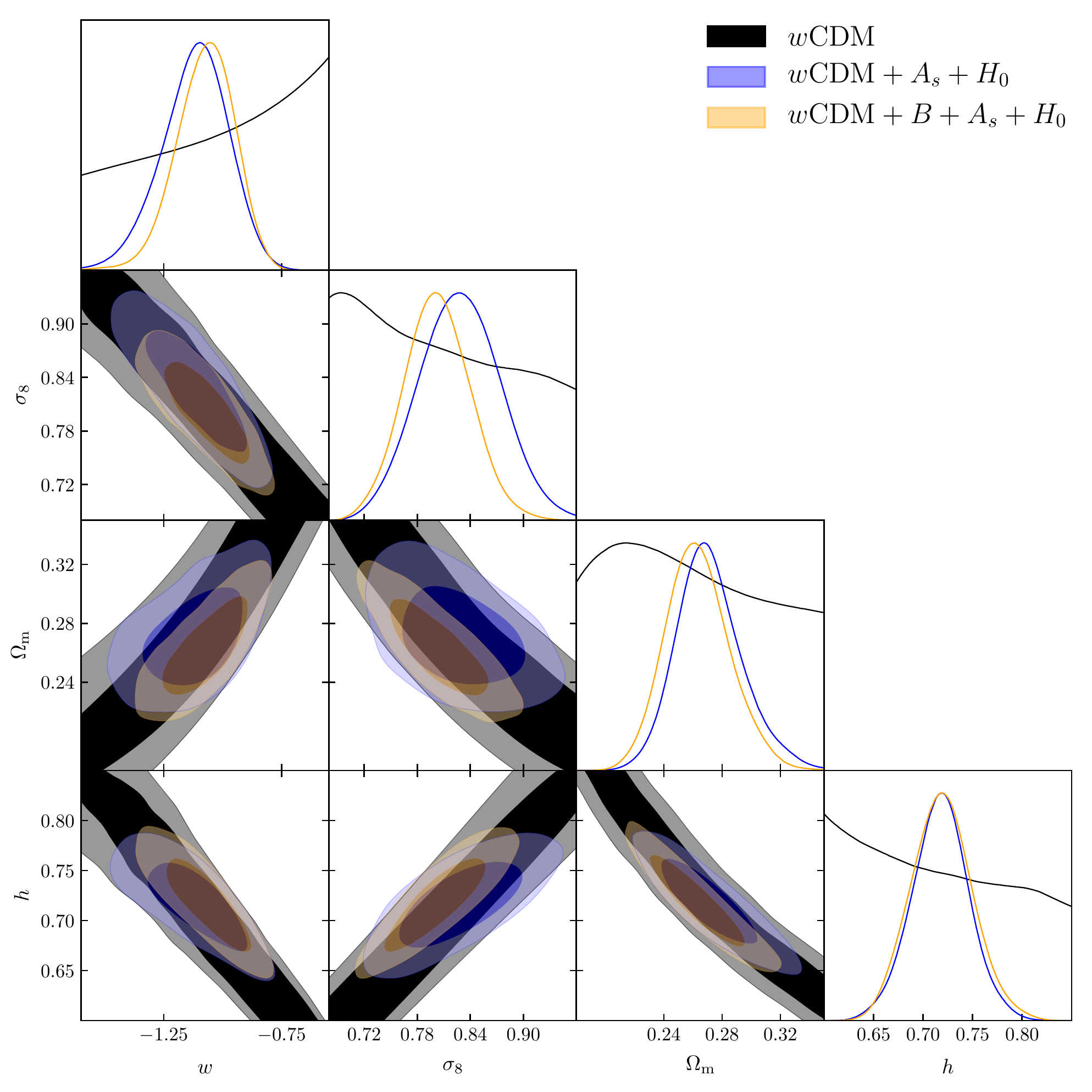}
\par\end{centering}
\caption{Marginalised (1D and 2D) joint posterior probability
 distributions of $w$, $\sigma_8$, $\Omega_\mathrm{m}$, and $h$. The black contours show the constraints from the tSZ power spectrum alone, while the blue contours show those from
 the tSZ combined with a local measurement of the Hubble constant, the Planck
 amplitude prior (see table \ref{tab:tll-1-1-1}), and $\tau_{\mathrm{reio}}=0.06\pm0.01$, for a flat prior on the mass bias. We fix the mass bias
 to $B=1.25$ for the yellow contours. \label{fig:tSZ+H}}
\end{figure*}
As dark energy slows down structure formation, we can constrain the dark
energy EoS $w$ by comparing the amplitude of fluctuations at $z=1090$
measured by the CMB and that in a low redshift universe measured by
tSZ. Specifically, the CMB constraint gives $\sigma_8$ as a function of $w$, or
$F$ as a function of $w$ and $B$; thus, we can trade $B$ for $w$.
In this section we constrain $w$ within the framework of a flat $w$CDM model.

The black contours in figure \ref{fig:tSZ+H} show marginalised joint
posterior probability distributions of $w$, $\sigma_8$,
$\Omega_\mathrm{m}$, and $h$ from the tSZ power spectrum {\it
alone}. As expected we cannot constrain individual parameters using the
tSZ power spectrum alone, but a constraint on the parameter $F$
(Eq.~\eqref{eq:F}) for $w$CDM, $F=0.460\pm0.013$ (68\%~CL), is similar
to that for $\Lambda$CDM.

To constrain $w$, we now add to our likelihood some external constraints
on the primordial scalar amplitude $A_s$ and $h$. The latter is needed
because $F$ contains $h$. We apply a Gaussian prior on $h=0.72\pm 0.03$
from local measurements of the Hubble constant \citep[see][and
references therein]{Bernal:2016gxb}. The CMB temperature anisotropy (without CMB lensing) data do not
constrain $A_s$, but only a combination $A_se^{-2\tau_{\mathrm{reio}}}$
because a small fraction of CMB photons from $z=1090$ are re-scattered
in a reionised universe at $z\lesssim 20$. While the exact constraint on
$A_se^{-2\tau_{\mathrm{reio}}}$ is somewhat model dependent, we follow
the procedure of the ``WMAP amplitude prior'' \citep{Komatsu:2008hk} to
find a robust Gaussian prior on
$10^9A_se^{-2\tau_{\mathrm{reio}}}=1.878\pm 0.014$ from the Planck 2015
data. See table~\ref{tab:tll-1-1-1} for summary. Finally, we use a 
Gaussian prior on the optical depth, motivated by the latest Planck measurement, i.e., $\tau_{\mathrm{reio}}=0.06\pm
0.01$ \citep[see][footnote 1]{Calabrese:2017ypx}, to obtain a prior on $A_s$. 

The mass bias also needs to be constrained before we can measure
$w$. With the flat prior given in
table~\ref{tab:Uniform-priors-imposed}, $1.11<B<1.67$, motivated by the scatter of results from numerical simulations, we find
$w=-1.15\pm0.15$ (68\%~CL). The blue contours in figure~\ref{fig:tSZ+H}
show 2d marginalised distributions. 
If we further assume a fixed value for the
mass bias, for instance we assume that {\it all} of the mass bias is due
to non-thermal pressure and take $B=1.25$ as in \cite{Dolag:2015dta}, we find 
\begin{equation}
w=-1.10\pm0.12 ~(68\%~\mathrm{CL}),\label{eq:tSZ}
\end{equation}
as well as $\sigma_{{8}}=0.802\pm0.037$ and
$\Omega_{{\mathrm{m}}}=0.265\pm0.022$. 

The yellow contours in
figure~\ref{fig:tSZ+H} show 2d marginalised distributions. In particular, the contours in the $h$-$w$ plane show a strong correlation. Hence, the central value of our constraint on the EoS depends primarily on the prior on $h$.  A smaller Hubble parameter would lead to a less negative $w$. 
\begin{table}
\begin{centering}
\begin{tabular}{lc}
 & $10^{9}A_{s}e^{-2\tau_{{\mathrm{reio}}}}$\tabularnewline
\hline 
$\Omega_{{\mathrm{k}}}=0$ and $w=-1$\dotfill & $1.880\pm0.014$\tabularnewline
$\Omega_{{\mathrm{k}}}\neq0$ and $w=-1$\dotfill & $1.872\pm0.014$\tabularnewline
$\Omega_{{\mathrm{k}}}=0$ and $w\neq-1$\dotfill & $1.880\pm0.014$\tabularnewline
$\Omega_{{\mathrm{k}}}=0$ and $w=-1$ and $m_{\nu}>0$ & $1.881\pm0.014$\tabularnewline
\hline 
Amplitude prior \dotfill & $1.878\pm0.014$\tabularnewline
\end{tabular}
\par\end{centering}
 \caption{Constraints on $10^{9}A_{s}e^{-2\tau_{{\mathrm{reio}}}}$
 at $k=0.05\mathrm{Mpc}^{-1}$ from the publicly available Planck 2015 chains assuming four
 different cosmological models. The last row shows our summary for the
amplitude prior.}\label{tab:tll-1-1-1}
\end{table}
\begin{table*}
\begin{centering}
\setcellgapes{2pt}\makegapedcells
\begin{tabular}{lc|ccccc}
 & 68\% C.L. &  $R$ & $\ell_{{\mathrm{A}}}$ & $\Omega_{{\mathrm{b}}}h^{2}$ & $10^{9}A_{{\mathrm{s}}}$ & $n_{{\mathrm{s}}}$ \tabularnewline
\hline 
$R$ & $1.7447\pm0.0068$ & $1$ & $0.51$ & $-0.66$ & $-0.58$ & $-0.85$\tabularnewline
$\ell_{{\mathrm{A}}}$ & $301.70\pm0.14$ & $0.51$ & $1$ & $-0.44$ & $-0.37$ & $-0.45$\tabularnewline
$\Omega_{{\mathrm{b}}}h^{2}$ & $0.02228\pm0.00024$ & $-0.66$ & $-0.44$ & $1$ & $0.39$ & $0.61$\tabularnewline
\hline
$10^{9}A_{{\mathrm{s}}}$ & $2.1108\pm0.0713$ & $-0.58$ & $-0.37$ & $0.39$ & $1$ & $0.60$\tabularnewline
$n_{{\mathrm{s}}}$ & $0.9681\pm0.0057$ & $-0.85$ & $-0.45$ & $0.61$ & $0.60$ & $1$\tabularnewline
\end{tabular}
\par\end{centering}
\caption{Posterior mean and covariance matrix derived from the Planck 2015
 ``lowTEB+lensing'' chains of $w$CDM. The second column gives the the mean
 values and 68\%CL standard deviations. The last five columns are the
 elements of cross-correlation coefficients $D_{ij}$.\label{tab:Compressed-likelihood,-base_w_pl}}
\end{table*}

How does this ``amplitude-derived'' constraint on $w$ compare with a
more common ``distance-derived'' constraint on $w$? To this end we form a compressed
likelihood (or a distance prior) of the distance information from CMB
data following the previous work
\citep{Mukherjee:2008kd,Komatsu:2008hk,Wang:2015tua}. The likelihood
includes the shift parameter $R$ and the angular scale of the sound
horizon at last scattering $\ell_{{\mathrm{A}}}$, as well as the baryon density. We emphasise that the compressed likelihood ignores the effect of dark energy on large scale perturbations, i.e., the late integrated Sachs-Wolfe effect, as it uses only the distance information. As a result, the dark energy constraint from the compressed likelihood is slightly weaker than the full analysis (though by not much). A comprehensive discussion on this point can be found in section 5.4 of \cite{Komatsu:2008hk}.

At each step of the MCMC, we compute
\begin{equation}
\chi_{{\mathrm{CMB}}}^{2}=\sum_{i,j=1}^{3}\left(P-\bar{P}\right)_{i}\left[C^{-1}\right]_{ij}\left(P-\bar{P}\right)_{j},
\end{equation}
where $P\equiv\left\{R,\ell_{{\mathrm{A}}}, \Omega_{{\mathrm{b}}}h^{2}\right\} $
contains the proposed values of the parameters and $\bar{P}$ contains
the posterior mean values of the $w$CDM ``lowTEB+lensing'' chains
\citep{Ade:2015xua}. The covariance matrix is given by
$C_{ij}=\sigma_{i}\sigma_{j}D_{ij}$. The posterior mean, the standard
deviation $\sigma_i$, and the correlation matrix $D_{ij}$ are given in the first three rows of table
\ref{tab:Compressed-likelihood,-base_w_pl}. To this we add a prior on
$h$ from the above, and form the total $\chi^2$ of
$\chi^{2}=\chi_{{\mathrm{CMB}}}^{2}+\chi_{\mathrm{Hubble}}^{2}$. We
find $w=-1.13\pm0.10$ and $\Omega_{{\mathrm{m}}}=0.274\pm0.023$
(68\%~CL), in good agreement with the tSZ+$H_{{0}}$ results. This shows
that the tSZ power spectrum is a compelling probe for dark energy,
complementary to distance-only constraints.

The predicted value of the matter fluctuation amplitude from the $w$CDM
CMB+$H_0$ is $\sigma_{{8}}=0.844\pm0.030$ (68\%~CL). We obtained this by
adding the information on the  primordial power spectrum to our
compressed likelihood (see last two rows of table
\ref{tab:Compressed-likelihood,-base_w_pl}). This is larger than what we
find from the tSZ+$H_0$ with $B=1.25$. Had we chosen a larger mass bias,
this apparent tension on $\sigma_8$ would be alleviated,
 at the expense of making $w$ more negative.

\section{Summary and conclusions}\label{sec:SC}
In this paper we have improved upon the calculation and likelihood
analysis of the tSZ power spectrum of resolved and
unresolved galaxy clusters and groups in a number of ways, and derived
a competitive constraint on the dark energy EoS parameter $w$ from the
amplitude of matter fluctuations.\\
\\

First we identified the source of differences in the analyses based on
the approach of \cite{Komatsu:2002wc} and that of the Planck
collaboration \citep{Ade:2013qta,Aghanim:2015eva}: it is due to
conversion of the virial mass to various over-density masses using some concentration-mass relations such as \cite{Duffy:2008pz}. Using HMF fits for
$M_{500\mathrm{c}}$ eliminates sensitivity to the choice of concentration-mass relations.\\

For the first time we incorporated all the important elements of the
likelihood analysis; namely, we include trispectrum in the tSZ power spectrum covariance matrix 
and vary the mass bias, the nuisance parameters, and all the relevant
cosmological parameters in the MCMC exploration of parameters. The
derived tSZ power spectrum with the nuisance parameters marginalised
over is significantly lower than that derived from the Planck 2015 analysis
\citep{Aghanim:2015eva}, which did not include trispectrum, at $\ell\gtrsim 300$.

We find that the tSZ power spectrum amplitude at $\ell\lesssim 10^3$
primarily depends on
$F=\sigma_{{{\mathrm{8}}}}\left(\Omega_{{\mathrm{m}}}/B\right)^{0.40}h^{-0.21}$. Using
the tSZ power spectrum data alone we find $F=0.460\pm 0.012$
(68\%~CL). In $\Lambda$CDM this implies a mass bias of $B=1.71\pm 0.17$
or $1-b=B^{-1}=0.58\pm 0.06$ (68\%~CL) when combined with the Planck CMB
data. This value agrees with that derived from the tSZ based cluster number counts 
\citep{Ade:2015fva}.

As dark energy slows down structure formation, we constrain $w$ by
combining the tSZ power spectrum, the primordial scalar amplitude
constrained by the Planck CMB data, and local Hubble constant
measurements. We find 
$w=-1.15\pm0.15$ (68\%~CL) for $1.11<B<1.67$ and $w=-1.10\pm 0.12$
(68\%~CL) for $B=1.25$. These constraints are consistent with, and
competitive and complementary to, more commonly studied distance-only
constraints on $w$ from CMB$+H_0$.

\section*{Acknowledgments}
We thank Ryu Makiya, Scott Kay, Jean-Baptiste Melin, Florian Ruppin,
Fabien Lacasa, Guillaume Hurier, Jens Chluba, Laura Salvati and James Colin Hill for discussions as well
as Thomas Tram and Julien Lesgourgues for their help with CLASS and
Montepython. We thank Antony Lewis for suggestions regarding the contour plots made the GetDist
software.  We thank the referee for useful comments and suggestions that helped improving the presentation and interpretation of our results. This work was supported in part by JSPS KAKENHI Grants,
JP15H05896 and ERC Consolidator Grant (CMBSPEC), No. 725456. This
analysis is based on observations obtained with Planck
(http://www.esa.int/Planck), an ESA science mission with instruments and
contributions directly funded by ESA Member States, NASA, and Canada.
\bibliographystyle{mnras} 
\bibliography{SZ_article_final}
\end{document}